\documentclass{gCOM2e}

\theoremstyle{plain}
\usepackage{xcolor}
\usepackage[font=small,labelsep=space]{caption}
\usepackage[hidelinks]{hyperref}

\usepackage{amsmath}

\usepackage[noautoalign, vcentermath, enableskew]{youngtab}\Yboxdim{12pt}

\usepackage{graphicx}
\usepackage{epsfig}
\usepackage{epstopdf}
\usepackage{framed}
\usepackage{longtable}

\newtheorem{thm}{Theorem}[section]
\newtheorem{prop}{Proposition}[section] 

\newtheorem{lem}{Lemma}[section]
\newtheorem{algo}{Algorithm}[section]
\usepackage[shortlabels]{enumitem}
\setlist[enumerate]{itemsep=0mm}

\theoremstyle{remark}
\theoremstyle{definition}

\newtheorem{defn}{Definition}[section]

\newcommand{\nten}{\mbox{10}}
\newcommand{\neleven}{\mbox{11}}
\newcommand{\ntwelve}{\mbox{12}}
\newcommand{\nthirteen}{\mbox{13}}
\newcommand{\nfourteen}{\mbox{14}}

\newcommand{\xone}{{\textrm{ }1}^{{ }^{{}^{\textrm{ }\textrm{ }1}}}}
\newcommand{\xtwo}{{\textrm{ }2}^{{ }^{{}^{\textrm{ }\textrm{ }2}}}}
\newcommand{\xthree}{{\textrm{ }3}^{{ }^{{}^{\textrm{ }\textrm{ }3}}}}
\newcommand{\xfour}{{\textrm{ }4}^{{ }^{{}^{\textrm{ }\textrm{ }5}}}}
\newcommand{\xfive}{{\textrm{ }5}^{{ }^{{}^{\textrm{ }\textrm{ }8}}}}
\newcommand{\xsix}{{\textrm{ }6}^{{ }^{{}^{\textrm{ }\textrm{ }4}}}}
\newcommand{\xseven}{{\textrm{ }7}^{{ }^{{}^{\textrm{ }\textrm{ }9}}}}
\newcommand{\xeight}{{\textrm{ }8}^{{ }^{{}^{\textrm{ }\textrm{ }6}}}}
\newcommand{\xnine}{{\textrm{ }9}^{{ }^{{}^{\textrm{ }\textrm{ }7}}}}
\newcommand{\xten}{{10}^{{ }^{{}^{12}}}}
\newcommand{\xeleven}{{11}^{{ }^{{}^{14}}}}
\newcommand{\xtwelve}{{12}^{{ }^{{}^{13}}}}
\newcommand{\xthirteen}{{13}^{{ }^{{}^{10}}}}
\newcommand{\xfourteen}{{14}^{{ }^{{}^{11}}}}

\begin{document}

\markboth{Dohan Kim}{International Journal of Parallel, Emergent and Distributed Systems}

\title{Representations of task assignments in distributed systems using Young tableaux and symmetric groups}

\author{Dohan Kim$^{\ast}$\thanks{$^\ast$ Email: dkim@airesearch.kr
\vspace{6pt}}  \\\vspace{6pt}  {\em{A.I. Research Co., 2537-1 Kyungwon Plaza 201, Sinheung-dong, Sujeong-gu, Seongnam-si, Kyunggi-do, 461-811, South Korea}}\\\vspace{6pt}
}

\maketitle

\begin{abstract}
This paper presents a novel approach to representing task assignments for partitioned agents (respectively, tasks) in distributed systems. A partition of agents (respectively, tasks) is represented by a Young tableau, which is one of the main tools in studying symmetric groups and combinatorics. In this paper we propose a task, agent, and assignment tableau in order to represent a task assignment for partitioned agents (respectively, tasks) in a distributed system. This paper is concerned with representations of task assignments rather than finding approximate or near optimal solutions for task assignments. A Young tableau approach allows us to raise the expressiveness of partitioned agents (respectively, tasks) and their task assignments.
\end{abstract}


\section{Introduction}
A distributed system is defined to be a collection of independent nodes that appear as a single coherent computer~\cite{Tanenbaum1995}. Parallel agents~\cite{Tompkins2003,Cosenza2011,Jang2005} in a distributed system often take advantage of parallelism~\cite{Tanenbaum1995,El-Rewini1998} by dividing a job into many tasks that execute on one or more agents. The primary purposes of task assignments in distributed systems are to increase the system throughput and to improve resource utilization~\cite{Wang1988,Efe1982,Shen1985,Lin1994,Shivaratri1992}. A subclass of task assignment problems involves an equal number of tasks and agents, where the mapping between a set of tasks and a set of agents is bijective. One of its fundamental form is represented by the \emph{linear assignment problem}~\cite{Burkard2009,Bus2002,Naiem2009,Cho1993}, which concerns how $n$ tasks are assigned to $n$ agents in the best possible way. Meanwhile, if tasks have a precedence relationship, they can be expressed as a directed acyclic graph (DAG), where each node represents a task and each edge represents a precedence constraint~\cite{Sinnen2007}. We focus on the representation of $n$-task-$n$-agent assignments for a given acyclic task graph $G=(V,\,E)$ in a distributed system. In our approach an $n$-task-$n$-agent assignment is represented by a \emph{Young tableau}~\cite{Fulton1997}. Our approach to representing an $n$-task-$n$-agent assignment is quite general, aiming to apply for task assignments involving the same number of tasks and agents in other disciplinary areas, such as robotics~\cite{Liu2012} and operations research~\cite{Burkard2009}. 

We also describe $n$-task-$m$-agent assignments $(n > m)$ by using \emph{generalized Young tableaux} and discuss their task reassignments by means of a group action on a set of generalized Young tableaux.

The remainder of this paper is organized as follows. We describe a task assignment problem including the $n$-task-$n$-agent assignment problem in a distributed system in Section~\ref{sec:TaskAssignmentProblem}. Section \ref{sec:Groups} provides an introduction to groups and Young tableaux. In Section~\ref{sec:symgroup} we discuss how an $n$-task-$n$-agent assignment can be represented by an element of a \emph{symmetric group} $\mathfrak{S}_n$~\cite{Sagan2001}. Section~\ref{sec:ScheduleTableaux} presents our approach to representing task assignments for agents in a distributed system using Young tableaux. We discuss an equivalence relation on a set of Young tableaux for $n$-task-$n$-agent assignments in this section. We also discuss generalized Young tableaux for $n$-task-$m$-agent assignments ($n>m$) and their equivalence relation on a set of generalized Young tableaux in this section. In Section~\ref{sec:taskreassignments} we discuss the counting aspect of the search space involving $n$-task-$n$-agent assignments and their reassignments by using \emph{tabloids} and symmetric groups~\cite{Sagan2001}. Finally, we conclude in Section~\ref{sec:Conclusions}.

\section{Task assignments in distributed systems}
\label{sec:TaskAssignmentProblem}

\subsection{Task assignment problem in a distributed system}
A task assignment problem in a distributed system is found in~\cite{Efe1982,Shen1985,Lin1994,Shahul2008} and is defined as follows:

Let $T$ be a set of $n$ tasks such that $T = \{t_1, t_2,\ldots, t_n\}$ and let $A$ be a set of $m\;(m \leq n)$ agents\footnote{In this paper we use \emph{agent} and \emph{processor} interchangeably if parallel agents in a distributed system are considered as simple computing entities~\cite{Tompkins2003}.} such that $A = \{a_1, a_2,\ldots, a_m\}$. Each task and agent is not necessarily homogeneous in a distributed system. A partial order relation $\prec$ can be defined on $T$, which specifies a task precedence constraint. For any two tasks $t_i, t_j \in T$, $t_i \prec t_j$  denotes that $t_i$ must be completed before $t_j$ can begin. Let \emph{M} be a task assignment between $T$ and $A$. Let $t_a^e(M)$ be the total execution time of agent $a$ for the task assignment $M$, and $t_a^i(M)$ be the total idle time of agent $a$ for the task assignment $M$. Agent $a$ is idle before the execution of its first task or between the executions of its two consecutive tasks for the task assignment $M$. The turnaround time of agent $a$ is the total time spent in agent $a$ for the task assignment $M$. Let $t_a (M) :=  t_a^e(M) + t_a^i(M) $ and  $t(M) := \max_{a}t_a (M)$, where $t(M)$ is called the \emph{task turnaround time} of the task assignment $M$. In contrast, let $u_a(M) := t_a^e(M)/t(M)$, where $u_a(M)$ is the agent utilization of agent $a$ for the task assignment $M$. The \emph{average agent utilization} for the task assignment $M$ is defined to be the average agent utilization for $m$ agents, i.e., $u(M) := (\sum_{k=1}^mu_{a_k}(M))/m$. If the performance metric for a task assignment is the task turnaround time, an optimal task assignment is defined to be the task assignment $M_0$ such that

\begin{center}
$t(M_0):=\displaystyle\min_Mt(M)=\displaystyle\min_M\displaystyle\max_at_a(M)$\;.
\end{center}

If the performance metric for a task assignment is the average agent utilization, an optimal task assignment is defined to be the task assignment $M_0$ such that
\begin{center}
$u(M_0):=\displaystyle\max_Mu(M)$\;.
\end{center}

Constraints and assumptions are as follows:
\begin{enumerate}[(1)]
\item Both tasks and distributed agents are not necessarily homogeneous and the information regarding their characteristics is available to a task assignment system. Agents are dedicated to a task assignment in which no other task or job is involved when a task assignment is executed.
\item The network of distributed agents is fully-connected in which communication links are identical with the same data transfer rate. Communications between agents take place by message passing.
\item Each task is assigned to exactly one agent in such a way that each agent is able to process only single task at a time in a non-preemptive manner. It is also required that at least one task is assigned to each agent.
\item Precedence constraints can be imposed. A task $t_j$ can be executed if all its predecessors $t_i$ with $t_i \prec t_j$ have completed. A task graph is directed and acyclic.
\end{enumerate}

Traditional approaches to representing task assignments have limitations in some cases. For instance, an assignment is often represented as a set of pairs (task ID, agent ID), graphs, matrices, charts, or tables~\cite{Burkard2009,Efe1982,Shen1985}. When we apply a logical partition to agents (or tasks) and their task assignments, those approaches often lack a systematic way of expressing a partition. Note that a partition in this paper refers to a logical partition of tasks or agents, which is different from the partition used in the context of \emph{grain packing}~\cite{Sinnen2007} that concerns how to partition a job into subtasks in order to improve the performance criteria. In our approach task assignments are represented by Young tableaux or tabloids in which partitions are naturally expressed. In Section~\ref{subsec:DefinitionsTerminologyforTaskAssignments} we provide the definitions and terminology for task assignments used in this paper.

\subsection{Definitions and terminology for task assignments}
\label{subsec:DefinitionsTerminologyforTaskAssignments}
In this subsection we introduce definitions and terminology for task assignments in a heterogeneous (agent-based) system. Definitions and terminology used in this subsection are found in~\cite{Wang1988,Suter2004,Sinnen2007,Benoit2008,El-Rewini1994,Ilavarasan2005,Shahul2008}.

A \emph{task graph} $T=(V,\,E)$ is a directed acyclic graph, where each node in $V=\{v_1, v_2,\ldots, v_n\}$ denotes a task, and each edge $(v_i,\,v_j) \in E \subset V \times V$ denotes a precedence relationship between tasks, i.e., $v_j$ cannot begin before $v_i$ completes. The positive weight associated with each task $v \in V$ represents a computation requirement. The nonnegative weight associated with each edge  $(v_i,\,v_j) \in E$ represents a communication requirement.

A fully-connected heterogeneous system $A$ is a set $A=\{a_1, a_2, \ldots, a_m\}$ of $m$ heterogeneous agents whose network topology is fully-connected. A heterogeneous system $A$ is called \emph{consistent} if agent $a_i \in A$ executes a task $n$-times faster than agent $a_j \in A$, then it executes all other tasks $n$-times faster than agent $a_j$. A heterogeneous system $A$ is called \emph{communication homogeneous} if each communication link in $A$ is identical.

Let $T=(V,\,E)$ be a task graph and $A=\{a_1, a_2, \ldots, a_m\}$ be a fully-connected heterogeneous  system. Assume that a startup cost of initiating a task on an agent is negligible. In a consistent system, the computation cost of task $v_i$ on $a_j$ is $\omega(v_i, a_j)=r(v_i)/e(a_j)$, where $r(v_i)$ is the computation requirement of task $v_i$, and $e(a_j)$ is the execution rate of agent $a_j$. Meanwhile, in an inconsistent system, the computation cost of task $v_i$ on $a_j$ is given by $\omega(v_i, a_j)=w_{ij}$, where $w_{ij}$ is the $(i, j)^\text{th}$ entry in a $|V| \times |A|$ cost matrix $W$. Note that an inconsistent system model is a generalization of a consistent system model. Now, the communication cost model of $A$ is defined as follows. Let $d(v_i, v_j)$ be the amount of data to be transferred from task $v_i $ to task $v_j$ for each $(v_i, v_j) \in E$; let $t(a_s, a_t)$ be the data transfer rate between the communication link between agent $a_s$ and $a_t$ in $A$. Let $M$ be a task assignment between $T$ and $A$ such that  $M(v_i)=a_s$ and $M(v_j)=a_t$ for $v_i, v_j \in T$ and $a_s, a_t \in A$. Assume that both local communication and communication startup cost are negligible. If the communication of $A$ has the linear cost model~\cite{Benoit2008}, the communication cost between task $v_i$ on agent $a_s$ and task $v_j$ on $a_t$ is given by $c(M(v_i), M(v_j))=d(v_i, v_j)/t(a_s, a_t)$ if $a_s \neq a_t$, and 0 otherwise. Furthermore, if $A$ is communication homogeneous such that the data transfer rate of each communication link is 1, the communication requirement and the communication cost coincide for inter-agent communication.

Let $T=(V,\,E)$ be a task graph and $A=\{a_1, a_2, \ldots, a_m\}$ be a fully-connected heterogeneous system. Suppose task $v_i \in V$ is assigned to agent $a_j \in A$, and a start time of task $v_i$ is $t_s(v_i, a_j)$. Then, the \emph{finish time} of task $v_i$ on agent $a_j$ is
\begin{center}
$t_f(v_i, a_j) := t_s(v_i, a_j) + \omega(v_i, a_j)$.
\end{center}

Let $T=(V,\,E)$ be a task graph and $A=\{a_1, a_2, \ldots, a_m\}$ be a fully-connected heterogeneous  system. The earliest possible start time of task $v_j \in V$ on agent $a_k \in A$ is called the \emph{data ready time},  which is defined as
\begin{center}
$t_{\text{dr}}(v_j, a_k) := \displaystyle\max_{v_i \in \text{pred}(v_j)}\{t_f(v_i, M(v_i))+c(M(v_i), M(v_j))\}$,
\end{center}
where $\text{pred}(v_j)$ denotes the set of all predecessors of task $v_j$, and $M(v_i)$ denotes the agent to which task $v_i$ is assigned by a task assignment $M$. If $\text{pred}(v_j) = \emptyset$, then $v_j$ is called an \emph{entry node}, and it is assumed that $t_{\text{dr}}(v_j, a_k) = 0$ for all $a_k \in A$.

\subsection{The $n$-task-$n$-agent assignment problem in a distributed system}
\label{defn:NtaskNprocessorAssignments}
\begin{figure}[h!]
  \centering
    \includegraphics[width=0.90\textwidth]{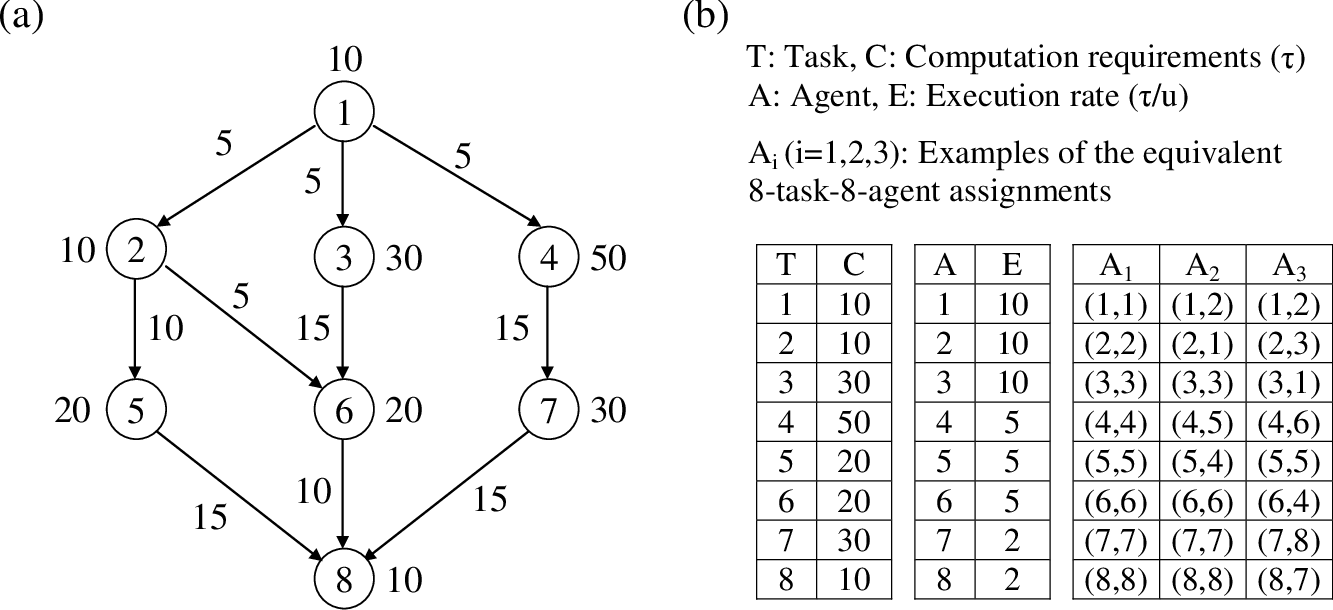}
  \caption{(a) Task graph $G=(V, E)$; (b) characteristics of tasks, agents, and examples of 8-task-8-agent assignments.}

\label{fig:TaskGraphAndAssignments}
\end{figure}

The $n$-task-$n$-agent assignment problem is a task assignment problem in a distributed system, which involves the same number of tasks and agents (cf. linear assignment problem~\cite{Burkard2009,Bus2002,Cho1993}). In the remainder of this paper, a target heterogeneous system $A$ for the $n$-task-$n$-agent assignment problem is assumed to be fully-connected, consistent, and communication homogeneous, where the communication requirement and the communication cost coincide. Figure~\ref{fig:TaskGraphAndAssignments} shows a task graph with eight tasks and examples of 8-task-8-agent assignments $A_i$ for $1\leq i \leq3$. The label next to each node in the task graph denotes the computation requirement and the label next to each edge denotes the communication requirement or communication cost. For instance, the communication cost between task 1 and 2 is 5 time units.

Consider the task assignment $A_1$ in Figure~\ref{fig:TaskGraphAndAssignments}(b). Each $(a, b)$ in $A_k$ for $1 \leq k \leq3$ denotes that task $a$ is assigned to agent $b$. Task 1 is the entry node in the task graph $G=(V, E)$, so it starts at time 0. Since task 1 is assigned to agent 1 in $A_1$, the computation cost of task 1 is its computation requirement divided by the execution rate of agent 1. A possible choice of units for $\tau$ and $u$ in Figure~\ref{fig:TaskGraphAndAssignments}(b) are Flop (Floating-point operation)~\cite{Tinetti1999} and second, respectively. Note that each task is assigned to each agent in such a way that their IDs are the same in $A_1$ (see Figure~\ref{fig:TaskGraphAndAssignments}(b)). Therefore, the computation cost of task 1 on agent 1 is $10/10 = 1$ time unit in $A_1$.  Since each task is assigned to a distinct agent for the $n$-task-$n$-agent assignment problem, each task starts at its data ready time. Thus, task 2 on agent 2 starts its execution at $1 + 5 = 6$ time units. Similarly, task 3 on agent 3 and task 4 on agent 4 start at 6 time units. Simple calculations show that task 5 on agent 5 starts at 17, task 6 on agent 6 at 24, task 7 on agent 7 at 31, and task 8 on agent 8 at 61 time units. Thus, the task turnaround time of  $A_1$ is $61 + 10/2 = 66$ time units, where 10/2 is a computation cost of task 8 on agent 8. Note that the execution rate of agents 1, 2, and 3 are the same (see Figure~\ref{fig:TaskGraphAndAssignments}(b)). Thus, it is indistinguishable in terms of the task turnaround time if we swap agents with the same execution rate in a given $n$-task-$n$-agent assignment. We see that the task turnaround time of $A_2$ and $A_3$ are the same with that of $A_1$. Furthermore, once the spatial assignment of tasks (i.e., allocation of tasks to agents~\cite{Sinnen2007}) has been determined, the temporal assignment of tasks (i.e., attribution of a start time to each task~\cite{Sinnen2007}) is deterministic for the $n$-task-$n$-agent assignment problem, i.e., the start time of each task on an agent is always its unique data ready time. Note also that if tasks in the $n$-task-$n$-agent assignment problem have no precedence relationship, the start time of each task on an agent is simply 0.

Traditional methods~\cite{Efe1982,Shen1985,Lin1994,Wang1988} to representing task assignments have some limitations if task assignments involve the same number of tasks and agents. If we apply an equivalence relation to tasks or agents, traditional approaches do not naturally express those assignments that belong to an equivalence class. We partition the search space $\mathfrak{S}_n$ of $n$-task-$n$-agent assignments by using an equivalence relation of Young tableaux of a given tableau shape. We introduce the necessary definitions and results of group theory and Young tableaux in Section~\ref{sec:Groups}.

\section{Groups and Young tableaux}
\label{sec:Groups}
Group theory is a branch of mathematics, which provides the methods, among other things, to analyze symmetry in both abstract and physical systems~\cite{Weisstein2003}. In this section we give definitions on groups and Young tableaux used in this paper. Definitions and results in this section are found in~\cite{Fraleigh1998,Hungerford1980,Alperin1995,Dummit2004,Sagan2001,Fulton1991,Fulton1997,Zhao2008,Kim2013}.

A \emph{group} $(G,\,\cdot\,)$ is a nonempty set \emph{G}, closed under a binary operation $\cdot$ , such that the following axioms are satisfied: (i) $(a\cdot b)\cdot c =  a \cdot (b \cdot c)$ for all $a,b,c \in G$, 
(ii) there is an identity element \emph{e} $\in$ \emph{G} such that for all $x\in G,~e \cdot x = x \cdot e = x$, (iii) for each element $a \in G$, there is an element $a^{-1} \in G$ such that $a \cdot a^{-1}= a^{-1} \cdot a = e$.

The \emph{order} of a finite group $G$, denoted $|G|$, is the number of elements of $G$.

Let $G$ be a group and $H$ be a nonempty subset of a group $G$. If $H$ itself is a group under the restriction to $H$ of the binary operation of $G$, then $H$ is a \emph{subgroup} of $G$, denoted by $H \leq G$. A subset $H$ of a group $G$ is a subgroup of $G$ iff (i) $H$ is closed under the binary operation of $G$, (ii) the identity $e$ of $G$ is in $H$, (iii) $h^{-1} \in H$ whenever $h \in H$. 

Let $I_n = \{1, 2,\ldots, n\}$. The group of all bijections $I_n  \rightarrow I_n$, whose binary operation is function composition, is called the \emph{symmetric group on n letters} and denoted $\mathfrak{S}_n$. Since $\mathfrak{S}_n$ is the group of all permutations of a set $I_n=\{1, 2,\ldots, n\}$, the order of $\mathfrak{S}_n$, i.e., $|\mathfrak{S}_n|$, is $n!$. A subgroup of a symmetric group is called a \emph{permutation group}. 

A permutation $\begin{pmatrix} 
1&2&\cdots&n\\
a_1&a_2 &\cdots&a_n
\end{pmatrix} \in \mathfrak{S}_n$ is written in the \emph{two-line notation}, while $a_1\,a_2\,\cdots\,a_n$$ \in \mathfrak{S}_n$ is written in the \emph{one-line notation}~\cite{Sagan2001}.

Let $i_1, i_2, \ldots, i_r\;(r \leq n)$ be distinct elements of $I_n = \{1, 2,\ldots, n\}$. Then $(i_1\,i_2\,\cdots\,i_r)$ is defined to be the permutation that maps $i_1 \mapsto i_2,\,i_2 \mapsto i_3,\,\ldots ,\, i_{r-1}\mapsto i_r$ and $i_r \mapsto i_1$, and every other element of $I_n$ maps onto itself. $(i_1\,i_2\,\cdots\,i_r)$ is called a cycle of length $r$ or an \emph{r-cycle}~\cite{Hungerford1980}. A 2-\emph{cycle} is called a \emph{transposition}~\cite{Hungerford1980,Fraleigh1998}. 

For instance, permutation $p=3\,1\,4\,2\in \mathfrak{S}_4$ is a 4-cycle such that $(1\,3\,4\,2)=(3\,4\,2\,1)=(4\,2\,1\,3)=(2\,1\,3\,4)$. 

Every permutation of a finite set can be written as a product of disjoint cycles. Any permutation of a finite set of at least of two elements can be written as a product of transpositions. If $p\in \mathfrak{S}_n$ is the product of disjoint cycles of lengths $l_1, l_2, \ldots, l_r$ with $l_1 \leq l_2 \leq \cdots \leq l_r$ (including its 1-cycles), the integers $l_1, l_2, \ldots, l_r$ are called the \emph{cycle type} of $p$. 

Let $G$ be a group and let $s_i \in G$ for $i \in I$. The subgroup generated by $S=\{s_i : i \in I\}$ is the smallest subgroup of $G$ containing the set $S$. If this subgroup is all of $G$, then $S$ is called a \emph{generating set} of $G$.

Let $G$ be a group. For any $H \leq G$ and any $g \in G$, let $gH=\{gh:h \in H\}$ and $Hg=\{hg: h\in H\}$. The former is called the \emph{left coset} of $H$ in $G$ and the latter is called the \emph{right coset} of $H$ in $G$.

Let $G$ be a group and let $H \leq G$. A subset $T=\{t_i\}$ of $G$ is called a \emph{(left) transversal} for $H$ in $G$ if the set $T$ consists of precisely one element from each left coset of $H$ in $G$. 

Let $G$ be a group and let $H \leq G$. The number of left cosets of $H$ in $G$ is called the \emph{index} $[G:H]$ of $H$ in $G$. If $G$ is a finite group, $[G:H]=|G|/|H|$.

Let $G$ be a group whose identity element is $e$. A (left) \emph{action} of $G$ on a set $X$ is a function $G \times X \rightarrow X$ such that for all $x \in X$ and $g_1, g_2 \in G$: (i) $ex = x$, (ii) $(g_1g_2)x =g_1(g_2x)$. When such an action is given, we say that $G$ acts (left) on the set $X$, and $X$ is called a $G$-set.

Let $X$ be a set. A relation $\sim_R$ on $X \times X$ is called an \emph{equivalence relation} on $X$ provided $\sim_R$ is: (i) reflexive: $x \sim_R x$ for all $x \in X$, (ii) symmetric: $x\sim_R y \Rightarrow y\sim_R x$, (iii) transitive: $x\sim_R y$ and $y\sim_R z  \Rightarrow x\sim_R z$. The \emph{equivalence class} of $x \in X$ under $\sim_R$ is defined to be the set $\{y \in X : y \sim_R x\}$.

Let $X$ be a $G$-set. For $x_1, x_2 \in X$, let $x_1 \sim_R x_2$ iff there exists $g \in G$ such that $gx_1 = x_2$. Then, $\sim_R$ is an equivalence relation on $X$. The equivalence classes with respect to $\sim_R$ are called the \emph{orbits} of $G$ on $X$.

A \emph{partition} of \emph{n} is defined to be a sequence $\lambda=(\lambda_1, \lambda_2,\ldots,\lambda_i)$, where the $\lambda_j$ are weakly decreasing and $\sum_{j=1}^i{\lambda_j}=n$. If $\lambda$ is a partition of \emph{n}, then it is denoted $\lambda \vdash n$.

Let $\lambda=(\lambda_1, \lambda_2, \ldots, \lambda_i) \vdash n$. A \emph{Young diagram} (or \emph{Ferrers diagram}) of shape $\lambda$ is a left-justified, finite collection of cells, or boxes, with row \emph{j} containing $\lambda_j$ cells for $1 \leq j \leq i$.

Let $\lambda \vdash n$. A \emph{Young tableau} of shape $\lambda$ is an array \emph{t}, obtained by assigning numbers in $\{1, 2,\ldots, n\}$ to the cells of the \emph{Young diagram} of shape $\lambda$ bijectively. 

Let $\lambda \vdash n$. A \emph{generalized Young tableau} of shape $\lambda$ is a filling of the \emph{Young diagram} of shape $\lambda$ with positive integers (repetitions allowed). 

For instance, let $\lambda=(2, 1)$. The list of all Young tableau of the shape $\lambda$ is as follows:
\begin{center}
$\young(12,3)$ \;,\; $\young(13,2)$ \;,\; $\young(21,3)$ \;,\; $\young(23,1)$ \;,\; $\young(31,2)$ \;,\; $\young(32,1)$ \;. 
\end{center}

Two Young tableaux $t_1, t_2$ of the same shape $\lambda$ are called \emph{row equivalent}, denoted $t_1 \sim t_2$, if the corresponding rows of $t_1$ and $t_2$ contain the same elements. (The reader is encouraged to verify that $\sim$ is an equivalence relation on the set of Young tableaux of shape $\lambda$.) A \emph{tabloid} of shape $\lambda$ is an equivalence class, defined as $\{t\}=\{t_1:t_1 \sim t\}$, where the shape of $t$ is $\lambda$.

To denote a tabloid $\{t\}$, only horizontal lines between rows are used. For instance,
\begin{center}
$t = \young(12,3)$\;\;\;implies\;\; $\{t\}=\left \{\young(12,3)\; , \; \young(21,3) \right \}=
\begin{tabular}{ccc}
\cline{1-2}\noalign{\smallskip}
1 & 2  \\
\cline{1-2}\noalign{\smallskip}
3  \\
\cline{1-1}\noalign{\smallskip}
\end{tabular}
\;.$ 
\end{center}

\begin{lem}[\cite{Sagan2001}]
\label{lem:Numbertabloid}
Suppose $\lambda = (\lambda_1, \lambda_2, \ldots, \lambda_i) \vdash n$. For a given tabloid of shape $\lambda$, the number of Young tableaux of shape $\lambda$ in the tabloid is $\lambda_1 !\lambda_2 ! \cdots \lambda_i !$. 
\end{lem}

\begin{lem}[\cite{Sagan2001}]
\label{lem:DimensionOfPermutationModule}
Let $\lambda = (\lambda_1, \lambda_2, \ldots, \lambda_i) \vdash n$. The number of distinct tabloids of shape $\lambda$ is $n!/(\lambda_1 ! \lambda_2 ! \cdots \lambda_i !).$
\end{lem}

Lemmas~\ref{lem:Numbertabloid} and~\ref{lem:DimensionOfPermutationModule} can be obtained by using simple combinatorial arguments. The interested reader may refer to~\cite{Sagan2001} for further details.

\section{Representations of $n$-task-$n$-agent assignments using a symmetric group}
\label{sec:symgroup}
Representations of bijective task assignments between tasks and agents (or processors) using a group theory have already been researched in~\cite{Kim2013,Rowe2002}. In this section we summarize how an $n$-task-$n$-agent assignment is represented by an element of $\mathfrak{S}_n$. 

Let $U_n$ be a set of $n$ tasks and $W_n$ be a set of $n$ distributed agents such that $U_n=W_n=\{1, 2,\ldots,n\}$. Then the group of all bijections $U_n \rightarrow W_n$ is $\mathfrak{S}_n$, where each element of $\mathfrak{S}_n$ denotes each $n$-task-$n$-agent assignment between $U_n$ and $W_n$. Therefore, a permutation 
$\begin{pmatrix} 
1&2&\cdots&n\\
a_1&a_2 &\cdots&a_n
\end{pmatrix}$
$\stackrel{\rm{def}}{=}a_1\,a_2\,\cdots\,a_n$$ \in \mathfrak{S}_n$~\cite{Sagan2001}
can be used to denote an $n$-task-$n$-agent assignment that maps $1 \mapsto a_1,2 \mapsto a_2, \ldots, n \mapsto a_n$, where $1, 2,\ldots,n \in U_n$ and $a_1, a_2,\ldots,a_n \in W_n$ such that $U_n=W_n=\{1, 2, \ldots,n\}$. 

Let $U_4=W_4=\{1, 2, 3, 4\}$. If permutation $p=3\,1\,4\,2=(1\,3\,4\,2)\in \mathfrak{S}_4$ is used to denote a $n$-task-$n$-agent assignment for $n=4$, then it can be represented by the set $\{(1\mapsto3), (2\mapsto1), (3\mapsto4), (4\mapsto 2)\}$, where $(a \mapsto b)$ denotes that task $a$ is assigned to agent $b$ for $a \in U_4$ and $b \in W_4$.

Although an $n$-task-$n$-agent assignment can be represented by the above manner, a task assignment for partitioned agents (or tasks) is not naturally represented. We will discuss this issue in the next section.

Meanwhile, a reassignment of an $n$-task-$n$-agent assignment can be represented by using permutation multiplication. For instance, if permutation $q=i_1\,i_2\,\cdots\,i_n \in \mathfrak{S}_n$ is used to denote an $n$-task-$n$-agent assignment, then the right multiplication of $q$ by transposition $(i\,j) \in \mathfrak{S}_n$ may represent the swapping of the task in agent $i$ and the task in agent $j$~\cite{Kim2013}. Permutation multiplication is further discussed in~\cite{Fraleigh1998,Kim2013}.

\section{Assignment tableaux and tabloids}
\label{sec:ScheduleTableaux}
\subsection{Assignment tableaux and tabloids for $n$-task-$n$-agent assignments}
\label{subsec:assigntabloids}
In this subsection we present our approach to representing $n$-task-$n$-agent assignments by using Young tableaux and tabloids. We first introduce a \emph{task tableau} and an \emph{agent tableau}. Then, we define an \emph{assignment tableau}, which is a 2-tuple of a task and agent tableau. 

\begin{defn}
Suppose $\lambda \vdash n$. A \emph{task tableau} of shape $\lambda$, denoted ${t}_{\lambda}$, is a Young tableau of shape $\lambda \vdash n$, obtained by assigning tasks (i.e., task IDs) in $\{1, 2,\ldots, n\}$ to the cells of the \emph{Young diagram} of shape $\lambda$ bijectively. An \emph{agent tableau} of shape $\lambda$, denoted ${a}_{\lambda}$, is a Young tableau of shape $\lambda$, obtained by assigning agents (i.e., agent IDs) in $\{1, 2,\ldots, n\}$ to the cells of the \emph{Young diagram} of shape $\lambda$ bijectively. 
\end{defn}

\begin{figure}[h!]
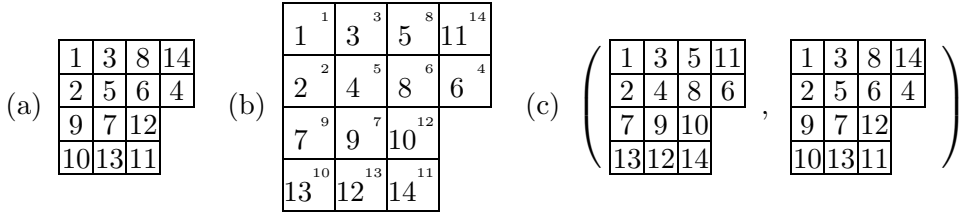

\begin{center}
(a)$\;\;\Yboxdim13pt\young(138\nfourteen,2564,97\ntwelve,\nten\nthirteen\neleven)$\;\;\;
(b)$\;\;\Yboxdim{20pt}\young(\xone\xthree\xfive\xeleven,\xtwo\xfour\xeight\xsix,\xseven\xnine\xten,\xthirteen\xtwelve\xfourteen)$\;\;\;
(c)$\;\; \left(\;\Yboxdim13pt\young(135\neleven,2486,79\nten,\nthirteen\ntwelve\nfourteen)\;\; , \;\;  \young(138\nfourteen,2564,97\ntwelve,\nten\nthirteen\neleven) \;\;\right)$\;
\end{center}
\caption{(a) An agent tableau; (b) a compact form of an assignment; (c) an assignment tableau.}
\label{fig:Assignmenttableaux}
\end{figure}

Suppose fourteen agents are partitioned into \{1, 3, 8, 14\}, \{2, 5, 6, 4\}, \{9, 7, 12\}, and \{10, 13, 11\}. It is naturally represented as an agent tableau in Figure~\ref{fig:Assignmenttableaux}(a). Suppose further that the 14-task-14-agent assignment is given by $\{(1\mapsto1), (3\mapsto3), (5\mapsto8), (11\mapsto14), (2\mapsto2), (4\mapsto5), (8\mapsto6), (6\mapsto4), (7\mapsto9), (9\mapsto7), (10\mapsto12), (13\mapsto10), (12\mapsto13), (14\mapsto11)\}$, where $(a \mapsto b)$ means task $a$ is assigned to agent $b$. This task assignment may have a compact form of a representation as shown in Figure~\ref{fig:Assignmenttableaux}(b), where the entry in each cell represents a task and the label in the upper right corner of each cell represents an agent. Since Figure~\ref{fig:Assignmenttableaux}(b) is not a standard form of a Young tableau, we describe this task assignment as a 2-tuple of Young tableaux instead. We use the task tableau of the same shape with that of the agent tableau as shown in Figure~\ref{fig:Assignmenttableaux}(c) in order to represent the task assignment corresponding to Figure~\ref{fig:Assignmenttableaux}(b). Now we define an \emph{assignment tableau} to represent an $n$-task-$n$-agent assignment.

\begin{defn}
An \emph{assignment tableau} of shape $\lambda$, denoted $s_{\lambda}$, is a 2-tuple of Young tableaux $s_{\lambda}\stackrel{\rm{def}}{=}({t}_{\lambda}, {a}_{\lambda})$, where $t_\lambda$ is a task tableau of shape $\lambda$ and $a_\lambda$ is an agent tableau of shape $\lambda$.
\end{defn}

An assignment tableau $s_{\lambda}$ represents a task assignment, where each task in a cell $(i, j)$ of $t_{\lambda}$ is assigned to each agent in a cell $(i, j)$ of $a_{\lambda}$ bijectively. Therefore, we also denote $s_\lambda$ as the set of all $(a \mapsto b)$~\cite{Knutson2008}, where $a$ is a task in a cell $(i, j)$ of $t_\lambda$ and $b$ is an agent in a cell $(i, j)$ of $a_\lambda$.

\begin{defn}
A \emph{standard agent tableau}\footnote{The reader is encouraged to compare the definition of a standard agent tableau with that of a \emph{standard Young tableau}~\cite{Sagan2001}. The latter is defined to be a Young tableau whose entries increase in each row and each column~\cite{Sagan2001}.} of shape $\lambda$, denoted $A_{\lambda}$, is an agent tableau having the entries of agent IDs $\{1, 2,\ldots, n\}$ in a sequential order, starting from the top left and ending at the bottom right. If an agent tableau is a standard agent tableau, then we say that the associated assignment tableau is \emph{standard}, denoted $S_{\lambda}$.
\end{defn}

\begin{figure}[h!]
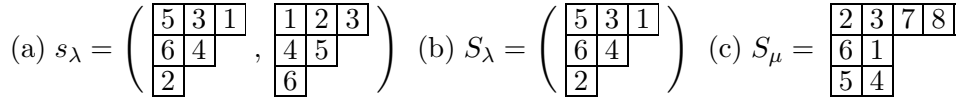

\begin{center}
(a)$\;s_{\lambda}=\left (\;\young(531,64,2) \; , \; \young(123,45,6) \;\right)\;$
(b)$\;S_{\lambda}=\left (\;\young(531,64,2)\;\right)\;$
(c)$\;S_{\mu}=\;\young(2378,61,54)$
\end{center}
\caption{(a) Assignment tableau; (b),(c) standard assignment tableaux.}
\label{fig:Stdassignmenttableaux}
\end{figure}

A simple renaming of agent IDs can be applied if necessary, in order to convert from an existing agent tableau of shape $\lambda \vdash n$ to the standard agent tableau of the same shape. Note that if an agent tableau is of shape $\lambda=(\lambda_1, \lambda_2,\ldots,\lambda_j)$, then the $\lambda_i (1\leq i \leq j)$ are weakly decreasing by the partition constraint of Young tableau. The following algorithm describes a simple renaming of agent IDs to convert from an existing agent tableau of shape $\lambda \vdash n$ to the standard agent tableau of the same shape. (It is exactly the same way to rename the task IDs in order to convert from an existing task tableau of shape $\lambda \vdash n$ to the standard task tableau of the same shape. Note that this renaming process has to be performed before task assignments.)

\begin{algo}\normalfont CONVERT-TABLEAU $(t_1, t_2, p)\mathrm{:}$\\
\setlist{nosep}
\noindent Input: an existing agent tableau $t_1$ of shape $\lambda \vdash n$, where  $\lambda=(\lambda_1, \lambda_2,\ldots,\lambda_j)$.\\
\noindent Output: the standard agent tableau $t_2$ of shape $\lambda \vdash n$, where  $\lambda=(\lambda_1, \lambda_2,\ldots,\lambda_j)$; permutation $p \in \mathfrak{S}_n$.
\begin{itemize}[leftmargin=*]
\item Write the entries of agent IDs in $t_1$ in the one-line permutation notation that is obtained by writing entries of $t_1$ in a sequential order starting from the top left and ending at the bottom right. Call the corresponding permutation for the obtained one-line permutation notation as $p \in \mathfrak{S}_n$.
\item Replace $t_1$ with the standard agent tableau $t_2$ of the same shape. Note that the corresponding permutation of $t_2$ for the one-line permutation notation is the identity permutation.
\item Return $t_2$ and permutation $p \in \mathfrak{S}_n$. (Permutation $p \in \mathfrak{S}_n$ might be used for the later purposes if the original agent IDs are not immaterial and need to be restored.)
\end{itemize}
\label{Algorithm:rmsending}
\end{algo}

Now, consider the assignment tableau $s_{\lambda} $ in Figure~\ref{fig:Stdassignmenttableaux}(a). Since the agent tableau in $s_{\lambda}$ is a standard agent tableau, it follows that $s_{\lambda}$ is a standard assignment tableau, i.e., $s_{\lambda}=S_{\lambda}$. Thus, we have $S_{\lambda} = s_\lambda $ $\stackrel{\rm{set}}{=} \{(5\mapsto1), (3\mapsto2), (1\mapsto3), (6\mapsto4), (4\mapsto5), (2\mapsto6)\}$ for $\lambda = (3,2,1)$. For a standard assignment tableau, we simply denote $S_{\lambda}$ as $(t_\lambda)$ rather than denoting a 2-tuple $(t_\lambda, a_\lambda)$. By a slight abuse of notation, if no confusion arises, we simply denote $S_{\lambda}$ as $t_\lambda$ without parentheses. For instance, Figure~\ref{fig:Stdassignmenttableaux}(c) represents a task assignment $S_{\mu}$ $\stackrel{\rm{set}}{=} \{(2\mapsto1), (3\mapsto2), (7\mapsto3), (8\mapsto4), (6\mapsto5), (1\mapsto6), (5\mapsto7), (4\mapsto8)\}$ for $\mu = (4,2,2)$, where eight agents are partitioned into three agent groups for $\mu=(4,2,2)$.

When we consider an assignment tableau, the partition constraints mandate that both task and agent tableau have the same shape. Once the agent tableau has chosen for an assignment tableau of shape $\lambda$, we may fix the agent tableau and consider the permutations of task tableaux of shape $\lambda$ in order to find other $n$-task-$n$-agent assignments. A standard assignment tableau is the preferred form for an $n$-task-$n$-agent assignment because a single tableau rather than a 2-tuple of Young tableaux represents an $n$-task-$n$-agent assignment between tasks and agents. 

We next describe a row-equivalence class of task, agent, and assignment tableaux.

\begin{defn}
\label{defn:AgentTabloid}
An \emph{agent tabloid} of shape $\lambda$, denoted $\{a_\lambda\}$, is a row-equivalence class of agent tableaux, i.e.,  $\{a_\lambda\} = \{a_{\lambda}' : a_{\lambda}' \sim a_\lambda\}$, such that agents in the same row of $a_\lambda$ have the same execution rate. A \emph{task tabloid} of shape $\lambda$, denoted $\{t_\lambda\}$, is a row-equivalence class of task tableaux, i.e., $\{t_\lambda\} = \{t_{\lambda}' : t_{\lambda}' \sim t_\lambda\}$.
\end{defn}

Agents that have the same execution rate are equivalent up to the $n$-task-$n$-agent assignment problem in terms of the task turnaround time. For instance, suppose that we have two distinct tasks $a$ and $b$, and two agents $x$ and $y$ that have the same execution rate. Then, 2-task-2-agent assignments $\{(a \mapsto x), (b \mapsto y)\}$, and $\{(a \mapsto y), (b \mapsto x)\}$ are equivalent in terms of the task turnaround time and the average agent utilization. In Figure~\ref{fig:TaskGraphAndAssignments}(b), the execution rates of agents in each set \{1,\,2,\,3\}, \{4,\,5,\,6\}, and \{7,\, 8\} are the same, respectively. In this case, we may represent them as an agent tabloid that has the entries of the first row 1, 2, and 3, the entries of the second row 4, 5, and 6, and the entries of the third row 7, and 8, respectively.

\begin{defn}
\label{defn:AssignmentTabloids}
An assignment tabloid of shape $\lambda$, denoted $\{s_{\lambda}\}$, is defined as a 2-tuple of $\{s_{\lambda}\} \stackrel{\rm{def}}{=} (t_\lambda, \{a_\lambda\})\stackrel{\rm{def}}{=}\{(t_\lambda, a_{\lambda}'):a_{\lambda}' \sim a_\lambda\}$. Equivalently, $\{s_{\lambda}\} \stackrel{\rm{def}}{=} (\{t_\lambda\}, a_\lambda)\stackrel{\rm{def}}{=}\{(t_{\lambda}', a_{\lambda}):t_{\lambda}' \sim t_\lambda\}$. If an agent tableau is a standard agent tableau or an agent tabloid containing a standard agent tableau, the associated assignment tabloid is said to be \emph{standard}, denoted $\{S_{\lambda}\}\stackrel{\rm{def}}{=}(\{t_\lambda\})$. By a slight abuse of notation, if no confusion arises, we also denote $\{S_{\lambda}\}$ as $\{t_{\lambda}\}$ (without parentheses).
\end{defn}

Since both definitions in Definition~\ref{defn:AssignmentTabloids} involve a row-equivalence class of Young tableaux of the same shape, the entry order in the same row within an assignment tableau is irrelevant, i.e., they are equivalent up to the $n$-task-$n$-agent assignment problem. In case an agent tabloid is given instead of an agent tableau, we replace the agent tabloid with the corresponding agent tableau, and the task tableau with the corresponding task tabloid in order to keep the canonical form of an assignment tabloid. For instance, if $t_\lambda = \young(31,2)$\;and\;$\{a_\lambda\}=
\begin{tabular}{ccc}
\cline{1-2}\noalign{\smallskip}
1 & 2  \\
\cline{1-2}\noalign{\smallskip}
3  \\
\cline{1-1}\noalign{\smallskip}
\end{tabular}
\text{, then }\{s_\lambda\}=\left(\,\young(31,2) \;,\;
\begin{tabular}{ccc}
\cline{1-2}\noalign{\smallskip}
1 & 2  \\
\cline{1-2}\noalign{\smallskip}
3  \\
\cline{1-1}\noalign{\smallskip}
\end{tabular}
\,\right).$ Equivalently, the above $\{s_\lambda\}$ can be written as $\{s_\lambda\}=\left (\,
\begin{tabular}{ccc}
\cline{1-2}\noalign{\smallskip}
1 & 3  \\
\cline{1-2}\noalign{\smallskip}
2  \\
\cline{1-1}\noalign{\smallskip}
\end{tabular}
\;,\; \young(12,3)\,\right)$. We see that $\{s_{\lambda}\}$ is a standard assignment tabloid. Thus,
$\{s_{\lambda}\}=\{S_{\lambda}\}=\;
\begin{tabular}{lcr}
\cline{1-2}\noalign{\smallskip}
1 & 3  \\
\cline{1-2}\noalign{\smallskip}
2  \\
\cline{1-1}\noalign{\smallskip}
\end{tabular}
$\;.

\begin{lem}
\label{lem:Numberofassignmentsbytabloid}
Suppose $\lambda = (\lambda_1, \lambda_2, \ldots, \lambda_i) \vdash n$. For a given assignment tabloid of shape $\lambda$, the number of $n$-task-$n$-agent assignments represented by the given assignment tabloid is $\lambda_1 !\lambda_2 ! \cdots \lambda_i !$. 
\end{lem}
\begin{proof}
Let $\{s_{\lambda}\}$ be an assignment tabloid such that $\{s_{\lambda}\}=(\{t_\lambda\}, a_\lambda)$. Then, the number of $n$-task-$n$-agent assignments represented by $\{s_{\lambda}\}$ for $\lambda \vdash n$ corresponds to the number of distinct task tableaux in $\{t_\lambda\}$. Therefore, the conclusion follows from Lemma~\ref{lem:Numbertabloid}.
\end{proof}

\begin{lem}
\label{lem:Numberofstandardassignmenttabloids}
Suppose $\lambda = (\lambda_1, \lambda_2, \ldots, \lambda_i) \vdash n$. Then, the number of distinct standard assignment tabloids of shape $\lambda$ is $n!/(\lambda_1 !\lambda_2 ! \cdots \lambda_i !)$.
\end{lem}
\begin{proof}
It immediately follows from Lemma~\ref{lem:DimensionOfPermutationModule}.
\end{proof}

For instance, consider standard assignment tabloids of the following shapes $\lambda_i \vdash n$ for $n=4$ and $1\leq i \leq3$: 
\begin{center}
$\lambda_1=(4): \yng(4)\;\;,\qquad$$\lambda_2=(1,1,1,1): \yng(1,1,1,1)\;\;, \qquad$$\lambda_3=(3,1): \yng(3,1)$\;\;.
\end{center}

The number of distinct standard assignment tabloids of the shape $\lambda_1$ is $(n! / \lambda_1 !)=(4! / 4!)=1$ for $n=4$. This situation may arise if all four agents are homogeneous. The number of distinct standard assignment tabloids of the shape $\lambda_2$ is $(n! / (1!)^n)=n!=4!$ for $n=4$. It corresponds to the number of all permutations of four tasks for a given standard agent tableau of shape $\lambda_2$. The number of distinct standard assignment tabloids of the shape $\lambda_3$ is $n! / (n-1)!=4$ for $n=4$, which corresponds to the number of choices (from 1 to 4) for the element in the second row.

\begin{prop}
\label{prop:EquivalentAssignmentsTableaux}
Let $\lambda = (\lambda_1, \lambda_2, \ldots, \lambda_i) \vdash n$. If $\{s_{\lambda}\}$ is an assignment tabloid such that $\{s_{\lambda}\}=(\{t_\lambda\}, a_\lambda)$, then $n$-task-$n$-agent assignments represented by $\{s_{\lambda}\}$ have the same task turnaround time (respectively, average agent utilization).
\end{prop}
\begin{proof}
By Definition~\ref{defn:AssignmentTabloids}, we have $\{s_{\lambda}\}=(\{t_\lambda\}, a_\lambda)=(t_\lambda, \{a_\lambda\})=\{(t_\lambda, a_{\lambda}'):a_{\lambda}' \sim a_\lambda\}$. Suppose to the contrary that the conclusion does not hold. Then, there exists two assignment tableaux $s^{1}_{\lambda}=(t_\lambda, a^{1}_{\lambda}) \in \{s_{\lambda}\}$ and $s^{2}_{\lambda}=(t_{\lambda}, a^{2}_{\lambda}) \in \{s_{\lambda}\}$ such that the task turnaround time (respectively, average agent utilization) of task assignments represented by $s^{1}_{\lambda}$ and $s^{2}_{\lambda}$ are not the same. It follows that there exists task $t$, and two agents $a$ and $b$ in the same row of $a^{1}_{\lambda}$ and $a^{2}_{\lambda}$, such that the task execution time of $t$ on $a$ and $t$ on $b$ necessarily differs, which is impossible by the choice of $a$ and $b$ since the execution rate of $a$ and $b$ are the same by Definition~\ref{defn:AgentTabloid}. Thus, we conclude that $n$-task-$n$-agent assignments represented by $\{s_{\lambda}\}$ have the same task turnaround time (respectively, average agent utilization).
\end{proof}

\begin{prop}
\label{prop:EquivalentStandardAssignmentsTableaux}
Let $\lambda = (\lambda_1, \lambda_2, \ldots, \lambda_i) \vdash n$. If $\{S_{\lambda}\}$ is a standard assignment tabloid such that $\{S_{\lambda}\}=\{t_\lambda\}$, then $n$-task-$n$-agent assignments represented by $\{S_{\lambda}\}$ have the same task turnaround time (respectively, average agent utilization).
\end{prop}
\begin{proof}
It immediately follows from Proposition~\ref{prop:EquivalentAssignmentsTableaux}.
\end{proof}

By Proposition~\ref{prop:EquivalentStandardAssignmentsTableaux}, the search space of the $n$-task-$n$-agent assignment problem can be reduced to the set of distinct standard assignment tabloids of a given shape $\lambda \vdash n$ instead of $\mathfrak{S}_n$. Note that the converse of Proposition~\ref{prop:EquivalentAssignmentsTableaux} and~\ref{prop:EquivalentStandardAssignmentsTableaux} is not necessarily true. Two $n$-task-$n$-agent assignments represented by two different assignment 
tabloids may have the same task turnaround time.

\begin{figure}[h!]
  \centering
	\includegraphics[width=0.90\textwidth]{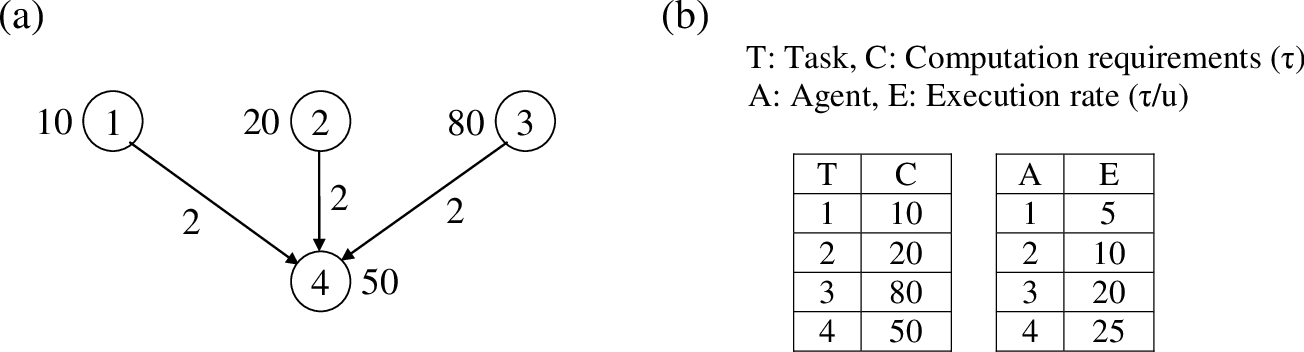}
  \caption{(a) Task graph; (b) characteristics of tasks, agents.}
\label{fig:DifferentAssignmentsWithSameOutput}
\end{figure}

Suppose we have four tasks and four heterogeneous agents as shown in Figure~\ref{fig:DifferentAssignmentsWithSameOutput}. Then, a standard assignment tabloid of shape $(1, 1, 1, 1)$ represents a $n$-task-$n$-agent assignment for $n=4$ in a unique manner. For instance, consider two different standard assignment tabloids of shape $(1, 1, 1, 1)$ having the task assignment sets $A_1 \stackrel{\rm{set}}{=}\{(1 \mapsto 1), (2\mapsto2), (3\mapsto3), (4\mapsto4)\}$ and $A_2\stackrel{\rm{set}}{=}\{(1\mapsto2), (2\mapsto1), (3\mapsto3), (4\mapsto4)\}$. A simple calculation shows that $A_1$ and $A_2$ have the same task turnaround time (8 time units) although their standard assignment tabloids are different.

Standard assignment tabloids are further studied in Section~\ref{sec:taskreassignments}, in which we consider the cases when tasks are reassigned for $n$-task-$n$-agent assignments represented by standard assignment tabloids of a given shape $\lambda \vdash n$.

\subsection{Generalized assignment tableaux for $n$-task-$m$-agent assignments}
\label{subsec:generalized}
An assignment tableau represents a task assignment, where each task in a cell $(i, j)$ of a task tableau is assigned to each agent in a cell $(i, j)$ of an agent tableau bijectively. If a set of tasks is assigned to a smaller-sized set of agents, i.e., $n$-task-$m$-agent assignment ($n > m$), we use generalized Young tableaux to represent $n$-task-$m$-agent assignments ($n > m$).

\begin{defn}
A \emph{generalized agent tableau} of shape $\lambda$, denoted $\bar{a}_{\lambda}$, is a filling of the \emph{Young diagram} of shape $\lambda$ with agents $\{1, 2,\ldots, n\}$ (repetitions allowed).
\end{defn}

\begin{defn}
A \emph{generalized assignment tableau} of shape $\lambda$, denoted $\bar{s}_{\lambda}$, is a 2-tuple of Young tableaux $\bar{s}_{\lambda}\stackrel{\rm{def}}{=}({t}_{\lambda}, \bar{a}_{\lambda})$, where $t_\lambda$ is a task tableau of shape $\lambda$ and $\bar{a}_\lambda$ is a generalized agent tableau of shape $\lambda$.
\end{defn}

\begin{defn}
\label{defn:stdgeneralizedtableau}
A \emph{standard task tableau} of shape $\lambda$, denoted $T_{\lambda}$, is a task tableau having the entries of task IDs $\{1, 2,\ldots, n\}$ in a sequential order, starting from the top left and ending at the bottom right. If a task tableau is the standard task tableau, then we say that the associated generalized assignment tableau is \emph{standard}, denoted $\bar{S}_{\lambda}$, i.e., $\bar{S}_{\lambda}\stackrel{\rm{def}}{=}({T}_{\lambda}, \bar{a}_{\lambda})$.
\end{defn}

\begin{figure}[h!]
\begin{center}
(a)$\;\bar{s}_{\lambda}=\left (\;\young(123,45,6) \; , \; \young(112,13,2) \;\right)\;$
(b)$\;\bar{S}_{\lambda}=\left (\;\young(112,13,2)\;\right)\;$
(c)$\;\bar{S}_{\lambda}=\;\young(112,13,2)$
\end{center}
\caption{(a) Generalized assignment tableau; (b),(c) standard generalized assignment tableaux.}
\label{fig:Generalizedassignmenttableaux}
\end{figure}

Consider the generalized assignment tableau $\bar{s}_{\lambda} $ in Figure~\ref{fig:Generalizedassignmenttableaux}(a). Since the task tableau in $\bar{s}_{\lambda}$ is a standard task tableau, $\bar{s}_{\lambda}$ is a standard generalized assignment tableau, i.e., $\bar{s}_{\lambda}=\bar{S}_{\lambda}$ by Definition~\ref{defn:stdgeneralizedtableau}. Thus, we have $\bar{s}_\lambda=\bar{S}_{\lambda}$ $\stackrel{\rm{set}}{=} \{(1\mapsto1), (2\mapsto1), (3\mapsto2), (4\mapsto1), (5\mapsto3), (6\mapsto2)\}$ for $\lambda = (3,2,1)$. As shown in Figure~\ref{fig:Generalizedassignmenttableaux}(b), we also denote $\bar{S}_{\lambda}$ as $(\bar{a}_\lambda)$ rather than denoting a 2-tuple $(t_\lambda, \bar{a}_\lambda)$ for a standard generalized assignment tableau. Similarly to a standard assignment tableau, by a slight abuse of notation, we may denote $\bar{S}_{\lambda}$ as $\bar{a}_\lambda$ without parentheses if no confusion arises (see Figure~\ref{fig:Generalizedassignmenttableaux}(c)). 

Note that a shape $\lambda$ of a standard assignment tableau $S_{\lambda}:=({t}_{\lambda}, {A}_{\lambda})$ (i.e., $S_{\lambda}:={t}_{\lambda}$) can be given based on a logical partition of agents. Meanwhile, a shape $\lambda$ of a standard generalized assignment tableau $\bar{S}_{\lambda}:=({T}_{\lambda}, {\bar{a}}_{\lambda})$ (i.e., $\bar{S}_{\lambda}:={\bar{a}}_{\lambda}$) can be given based on a logical partition of tasks rather than agents. Unlike $n$-task-$n$-agent assignments, a standard generalized assignment tableau alone does not necessarily determine the task turnaround time of a task assignment. As mentioned in Section~\ref{defn:NtaskNprocessorAssignments}, the start time of each task for an $n$-task-$n$-agent assignment is its unique data ready time. Meanwhile, the start time of each task for an $n$-task-$m$-agent assignment for $n > m$ depends on an execution order, i.e., temporal assignment of tasks. For instance, if agent $a$ has two tasks $t_1$ and $t_2$ with the same data ready time, then it depends on a task assignment algorithm to determine whether $t_1$ or $t_2$ starts first on $a$. However, if no task precedence constraint is given, we may obtain the task turnaround time of a given standard generalized assignment tableau using the $|T| \times |A|$ cost matrix, where $|T|$ is the number of tasks and $|A|$ is the number of agents.

In Section~\ref{subsec:assigntabloids} we showed that the search space of the $n$-task-$n$-agent assignment problem can be reduced to the set of standard assignment tabloids of a given shape $\lambda \vdash n$. We now consider a search space for the $n$-task-$m$-agent assignment problem ($n > m$) consisting of standard generalized assignment tableaux of a given shape. Let $X_j^\lambda=\{t_1, t_2, \ldots,t_j\}$ denote a set of distinct standard generalized assignment tableaux $t_i (1\leq i \leq j)$ of a given shape $\lambda$. If $j=n^m$, then $X_j^\lambda$ is the whole search space of the $n$-task-$m$-agent assignment problem represented by standard generalized assignment tableaux of a given shape $\lambda$ (see the following ``Remarks''). If $j = k$ for $k < n^m$, it is a selected search space of the $n$-task-$m$-agent assignment problem represented by standard generalized assignment tableaux of a given shape $\lambda$.\\

\noindent{\textbf{Remarks. } The number of distinct standard general assignment tableaux of shape $\lambda= (\lambda_1, \lambda_2, \ldots, \lambda_i) \vdash n$ for $m$ agents is $n^m$. Using an elementary counting argument, we see that the number of distinct standard generalized assignment tableaux of shape $\lambda \vdash n$ for $m$ agents is the same with the number of $m$ permutations from the set of $n$ tasks with repetition allowed, which is $n^m$. Note that it does not rely on a given shape of standard generalized assignment tableaux. }\\

The following proposition describes an equivalence relation on a set of standard generalized assignment tableaux of a given shape with respect to task reassignments defined by a group action. 
\begin{prop}
\label{prop:orbit}
Let $G \leq \mathfrak{S}_n$ act on $X_n^\lambda=\{t_1, t_2,\ldots,t_n\}$ by $gt_i=t_{g(i)}$ for $g \in G$. For  $t_1,t_2 \in X_n^\lambda$, let $t_1 \sim_{tr} t_2$ iff there exists $g \in G$ such that $gt_1=t_2$. Then, $\sim_{tr}$ is an equivalence relation on $X_n^\lambda$.
\end{prop}
\begin{proof}
Since $et=t$ for each $t\in X_n^\lambda$, we have $t\sim_{tr} t$. Thus, $\sim_{tr}$ is reflexive.\\
\indent To show that $\sim_{tr}$ is symmetric, assume $t_1 \sim_{tr} t_2$ for $t_1, t_2 \in X_n^\lambda$. It follows that $gt_1=t_2$ for some $g\in G$. Now, we have $g^{-1}(gt_1)=g^{-1}(t_2)$. Since $g^{-1}(t_2)=g^{-1}(gt_1)=et_1=t_1$, we have $t_2 \sim_{tr} t_1$. Thus, $\sim_{tr}$ is symmetric.\\
\indent To show that $\sim_{tr}$ is transitive, assume that $t_1 \sim_{tr} t_2$ and $t_2 \sim_{tr} t_3$ for $t_1, t_2, t_3 \in X_n^\lambda$. Then, we have $g_1t_1=t_2$ and $g_2t_2=t_3$ for some $g_1, g_2 \in G$. We claim that $(g_2g_1)t_1=t_3$, which shows that $t_1 \sim_{tr} t_3$. Since $(g_2g_1)t_1 = g_2(g_1t_1)=g_2(t_2)=t_3$, we have $(g_2g_1)t_1=t_3$. Thus, $t_1\sim_{tr} t_3$, which shows that $\sim_{tr}$ is transitive.
\end{proof}

The following proposition describes the size of the equivalence class of $t \in X_n^\lambda$ with respect to the equivalence relation $\sim_{tr}$. We first denote the equivalence class of $t \in X_n^\lambda$ with respect to $\sim_{tr}$ by $G(t)=\{gt: g\in G\}$ when $G \leq \mathfrak{S}_n$ act on $X_n^\lambda=\{t_1, t_2, \ldots,t_n\}$ by $gt_i=t_{g(i)}$ for $g \in G$.

\begin{prop}
\label{prop:sizeoforbit}
Let $G \leq \mathfrak{S}_n$ act on $X_n^\lambda=\{t_1, t_2, \ldots,t_n\}$ as above and let ${G}^t =\{g\in G:gt=t\}$ for $t \in X_n^\lambda$. The size of the equivalence class of $t \in X_n^\lambda$ with respect to the equivalence relation $\sim_{tr}$ in Proposition~\ref{prop:orbit} is $|G|/|{G}^t|$. 
\end{prop}
\begin{proof}
We show that $g{G}^t\mapsto gt$ is a well-defined bijection from the set of cosets of ${G}^t$ in $G$ onto the equivalence class $G(t)=\{gt: g\in G\}$ for $t \in X_n^\lambda$. We first show that ${G}^t$ is a subgroup of $G$.
\\
\indent Let $a, b \in G^t$. Then, we have $a t = t$ and $b t= t$. It follows that $(ab)t=a(bt)=at=t$. Thus, $ab \in G^t$, which shows that $G^t$ is closed under the binary operation of $G$. Since $et=t$, we have $e \in G^t$ as well. Let $h \in G^t$. Then, we have $ht=t$. Since $t=et=(h^{-1}h)t=h^{-1}h(t)=h^{-1}t$, it follows that $h^{-1} \in G^t$. Thus, $G^t$ is a subgroup of $G$.\\
\indent Let $g_1,g_2 \in G$. Since $g_1t=g_2t \Longleftrightarrow g_2^{-1}g_1t=t \Longleftrightarrow g_2^{-1}g_1 \in {G}^t \Longleftrightarrow g_1{G}^t=g_2{G}^t$, we see that $g{G}^t \mapsto gt$ is a well-defined bijection.\\
\indent Thus, the size of the equivalence class of $t \in X_n^\lambda$ with respect to the equivalence relation $\sim_{tr}$ in Proposition~\ref{prop:orbit} is $[G:G^t]=|G|/|{G}^t|$.
\end{proof}

Proposition~\ref{prop:sizeoforbit} directly follows from the \emph{orbit-stabilizer theorem}~\cite{Holt2005} in group theory. The interested reader may refer to~\cite{Fraleigh1998,Hungerford1980,Holt2005} for further details.\\
\indent Now, we illustrate how Proposition~\ref{prop:sizeoforbit} applies to a simple search space consisting of standard generalized assignment tableaux of a given shape. We assume that task reassignments are closed with respect to a search space, i.e., if a task reassignment transforms a standard generalized assignment tableau $t_i$ to $t_j$, then both $t_i$ and $t_j$ belong to the given search space consisting of standard generalized assignment tableaux of a given shape. \\
\indent For instance, let $G=\{e, (2\,3\,5\,6), (2\,5)(3\,6), (2\,6\,5\,3)\}$ be a subgroup of $\mathfrak{S}_8$ and let $X_8^\lambda=\{t_1, t_2, \ldots,t_8\}$ be a set of distinct standard generalized assignment tableaux of a given shape $\lambda$. Let $G$ act on $X_8$ as above.  Since ${G}^{t_2}={G}^{t_3}={G}^{t_5}={G}^{t_6}=\{e\}$, the size of the equivalence class of each $t_2, t_3, t_5$, and $t_6$ is 4 by Proposition~\ref{prop:sizeoforbit}. Similarly, ${G}^{t_1}={G}^{t_4}={G}^{t_7}={G}^{t_8}=G$. It follows that the size of the equivalence class of each $t_1, t_4, t_7$, and $t_8$ is $[G:G]=1$. \\
\indent In contrast, the reader is encouraged to verify that if $\mathfrak{S}_8$ acts on $X_8^\lambda$ as above, the equivalence class of each $t_i$ for $1\leq i \leq 8$ is the whole $X_8^\lambda$.\\
Now, we illustrate an algorithm for finding the equivalence class of a standard generalized assignment tableau of a given shape with respect to $\sim_{tr}$. The following algorithms are slight modifications of the \emph{orbit-stabilizer algorithms} discussed in~\cite{Holt2005, Hulpke2010}\footnote{In computational group theory right group actions are often considered~\cite{Holt2005, Hulpke2010} for convenience. However, we are only concerned with left group actions and stick with them throughout this paper.}. \\
\indent We first convert $X_n^\lambda$ into a totally ordered set $({X_n^\lambda}, \leq_t)$ by assigning each $t_i \in X_n^\lambda$ to a number, denoted $|t_i|$, which is the sum of all entries of $t_i$~\cite{Prosper2000}. The total order $\leq_t$ is defined on $X_n^\lambda$ in such a way that $t_i \leq_t t_j$ iff $|t_i| \leq |t_j|$. Then, we rename each tableau of $X_n^\lambda$ in such a manner that $t_1 \leq_t t_2\leq_t\cdots\leq_t t_n$. 
\\
\begin{algo}
\label{algo:orbit}
\normalfont TASK-REASSIGNMENT-ORBIT $(t_j, {X_n^\lambda}, \Omega)\mathrm{:}$\\
\setlist{nosep}
\noindent \textbf{Input:} a standard generalized assignment tableau $t_j \in {X_n^\lambda}$ of shape $\lambda \vdash k$, a permutation group $G \leq \mathfrak{S}_n$ given by a generating set $\Omega=\{g_1, \ldots, g_m\}$.\\
\noindent \textbf{Output:} the equivalence class of a given standard generalized assignment tableau $t_j$ with respect to $\sim_{tr}$.\\\\
1\;\;\;\;\;$\Delta:=[t_j]$;\\
2\;\;\;\;\;\textbf{for }$\alpha \in \Delta,\,\, g\in \Omega$ \textbf{do  }\\
3\;\;\;\;\;\;\;\;\;\;\ $\beta=g\alpha$;\\ 
3\;\;\;\;\;\;\;\;\;\;\textbf{if } $\beta \notin \Delta$\;\;\;\;\;\textbf{then}\\ 
4\;\;\;\;\;\;\;\;\;\;\;\;\;\;\; append $\beta$ to $\Delta$;\\
5\;\;\;\;\;\textbf{return} $\Delta$;
\label{Algorithm:rmsending}
\end{algo}
$\\$
In Algorithm~\ref{algo:orbit} a permutation group is given by the set of $m$ generators. Therefore, there will be $|\Delta|m$ images to compute.\\
\indent As an example of Algorithm~\ref{algo:orbit}, consider $({X_{50}^\lambda}, <_t)$, where $X_{50}^\lambda=\{t_1, t_2, \ldots,t_{50}\}$ and $t_1 <_t t_2 <_t \cdots <_t t_{50}$ for standard generalized assignment tableaux $t_i (1 \leq i \leq 50)$ of shape $\lambda \vdash k$. (We write $t_i<_t t_j$ if $t_i \leq_t t_j$ and $|t_i| \neq |t_j|$.) If $G\leq \mathfrak{S}_{50}$ act on $X_{50}^\lambda=\{t_1, t_2, \ldots,t_{50}\}$ by $gt_i=t_{g(i)}$ for $g \in G$ as discussed and a group $G$ is generated by $\Omega=\{(1\,2), (2\,3), (3\,4)\}$, then the equivalence class of $t_1$ with respect to $\sim_{tr}$ is $\{t_1, t_2, t_3, t_4\}$. On the other hand, if a group is generated by $\Omega^\prime=\{(47\,48), (48\,49), (49\,50)\}$, then the equivalence class of $t_{48}$ with respect to $\sim_{tr}$ is $\{t_{47}, t_{48}, t_{49}, t_{50}\}$. A typical example of a standard generalized assignment tableau in $\{t_1, t_2, t_3, t_4\}$ is one with smaller number entries from $\{1, 2, \ldots, k\}$, while a typical example of a standard generalized assignment tableau in $\{t_{47}, t_{48}, t_{49}, t_{50}\}$ is one with larger number entries from $\{1, 2, \ldots, k\}$.\\
\indent Let $G(t_j)$ denote the equivalence class of $t_j \in X_n^\lambda$ with respect to $\sim_{tr}$. For each $\sigma \in G(t_j)$, the following algorithm finds an element $g \in G$ such that $gt_j=\sigma$. It returns two lists $\Delta$ and $L$, in which $\Delta$ stores each $\sigma \in G(t_j)$, while $L$ stores the corresponding group element $g \in G$ satisfying $gt_j=\sigma$. The first elements of $\Delta$ and $L$ are $t_j$ and $e$, respectively. In Algorithm~\ref{algo:transversal} $L[\sigma]$ denotes $L[i]$ where $\Delta[i]=\sigma$. Note that if we have two arbitrary $\sigma_1, \sigma_2 \in G(t_j)$, we can find $g\in G$ such that $g\sigma_1=\sigma_2$ using the following algorithm by obtaining $g_1t_j = \sigma_1$ and $g_2t_j=\sigma_2$ (i.e., $g=g_2{g_1}^{-1}$).
$\\$
\begin{algo}
\label{algo:transversal}
\normalfont TASK-REASSIGNMENT-TRANSVERSAL $(t_j, {X_n^\lambda}, \Omega)\mathrm{:}$\\
\setlist{nosep}
\noindent \textbf{Input:} a standard generalized assignment tableau $t_j \in {X_n^\lambda}$ of shape $\lambda \vdash k$, a permutation group $G \leq \mathfrak{S}_n$ given by a generating set $\Omega=\{g_1, \ldots, g_m\}$.\\
\noindent \textbf{Output:} Transversal $L$, the equivalence class of a given standard generalized assignment tableau $t_j$ with respect to $\sim_{tr}$.\\\\
1\;\;\;\;\;$\Delta:=[t_j]$;\\
2\;\;\;\;\;$L:=[e]$;\\
3\;\;\;\;\;\textbf{for }$\alpha \in \Delta$ \textbf{do  }\\
4\;\;\;\;\;\;\;\;\;\;\textbf{for }$i \in \{1, \ldots, m\}$ \textbf{do  }\\
5\;\;\;\;\;\;\;\;\;\;\;\;\;\;\;$\beta=g_i\alpha$;\\
6\;\;\;\;\;\;\;\;\;\;\;\;\;\;\;\;\;\;\;\;\textbf{if } $\beta \notin \Delta$\;\;\textbf{then}\\ 
7\;\;\;\;\;\;\;\;\;\;\;\;\;\;\;\;\;\;\;\;\;\;\;\;\;\;\;\;\;\; append $\beta$ to $\Delta$;\\
8\;\;\;\;\;\;\;\;\;\;\;\;\;\;\;\;\;\;\;\;\;\;\;\;\;\;\;\;\;\; append $g_i\cdot L[\alpha]$ to $L$;\\
9\;\;\;\;\;\textbf{return} $L$, $\Delta$;
\end{algo}
$\\$
\indent For instance, consider when $G\leq \mathfrak{S}_{50}$ acts on $X_{50}^\lambda=\{t_1, t_2, \ldots,t_{50}\}$ by $gt_i=t_{g(i)}$ for $g \in G$ and a group $G$ is generated by $\Omega=\{g_1, g_2, g_3\}$, where $g_1=(1\,2)$, $g_2=(2\,3)$, and $g_3=(3\,4)$. As seen above, the equivalence class of $t_1$ (i.e., $G(t_1))$ with respect to $\sim_{tr}$ is $\{t_1, t_2, t_3, t_4\}$. Now, we find a group element $g \in G$ such that $gt_1 = t_4$ using Algorithm \ref{algo:transversal}. The first elements of $\Delta$ and $L$ are given as $t_1$ and $e$, respectively (i.e., $\Delta[1]=t_1$ and $L[1]=e$). The reader is encouraged to verify that $\Delta[2]=t_2$, $L[2]=(1\,2)$, $\Delta[3]=t_3$, $L[3]=(2\,3)(1\,2)$, $\Delta[4]=t_4$, and $L[4]=(3\,4)(2\,3)(1\,2)$. Since $\Delta[4]=t_4$, we have $g=L[4]=(3\,4)(2\,3)(1\,2)$. \\
\indent The time-complexity of both Algorithm~\ref{algo:orbit} and Algorithm~\ref{algo:transversal} is proportional to $|\Delta|m$. To test whether $\beta \in \Delta$ or $\beta \notin \Delta$ for both algorithms is basically a search problem. An additional data structure (e.g., list) can be used for $\Delta$ to maintain the sorted hash value for each entry $t_i$ in $\Delta$.\\
\indent The correctness of Algorithms~\ref{algo:orbit} and \ref{algo:transversal} directly follows from the \emph{orbit algorithms}~\cite{Holt2005, Hulpke2010} used in GAP~\cite{GAP2014} by renaming the elements of $X_n^\lambda$ such that $t_i \mapsto i$ for $t_i \in X_n^\lambda$ and defining the action of $G\leq \mathfrak{S}_n$ on the resulting set, denoted $\overline{X}_n^\lambda=\{1, 2, \ldots, n\}$, simply by $gi=g(i)$ for $g \in G, i \in \overline{X}_n^\lambda$ accordingly. The total order $\leq$ is naturally defined on the set $\overline{X}_n^\lambda$. (The orbit algorithms have already been well-established when $G\leq \mathfrak{S}_n$ acts on the set $\{1, 2, \ldots, n\}$ as above~\cite{GAP2014}. In Appendix B we discuss how GAP is used for the applications of Propositions~\ref{prop:orbit} and~\ref{prop:sizeoforbit}.) \\
\indent Note that the total order defined on $X_n^\lambda$ in this section is not the only total order that can be defined on $X_n^\lambda$. The purpose of defining an equivalence relation and a total order on $X_n^\lambda$ is to explore the search space corresponding to $X_n^\lambda$ in a well-defined and systematic manner by means of algebraic operations (e.g., a group action). We leave it as an open question to consider and define other useful total orders on $X_n^\lambda$ that exploits the symmetry during the navigation of a search space.

\section{Task reassignments on the set of standard assignment tabloids of shape $\lambda \vdash n$}
\label{sec:taskreassignments}
The counting aspects\footnote{The counting argument of task assignments in distributed systems was researched in~\cite{Shin1990}. However, its scope is quite different from that of our approach.} of the search space of the $n$-task-$n$-agent assignment problem using tabloids of a given shape are further investigated in this section. 

We first give a brief overview of the necessary background of this section. We assume that the reader has some familiarity with a \emph{vector space}~\cite{Fraleigh1998} over a \emph{field} $\mathbb{K}$~\cite{Dummit2004}. Examples of fields are the set of real numbers $\mathbb{R}$ and the set of complex numbers $\mathbb{C}$. In the remainder of this section, $G$ denotes a finite group, $\mathbb{K}$ a field, and $V$ denotes a finite dimensional vector space. Definitions and results used in the following overview are found in~\cite{Fraleigh1998,Alperin1995,Kim2013,Rotman1965,Sagan2001,Dummit2004,Hungerford1980,Fulton1991}.

Suppose $G$ acts on $V$ over $\mathbb{K}$. The action of $G$ on $V$ is called \emph{linear} if the following conditions are met: (i) $g(v+w)=gv+gw$ for all $g \in G$ and $v, w \in V$,
(ii) $g(kv)=k(gv)$ for all $g \in G, k \in \mathbb{K}$, and $v \in V$. If $G$ acts on $V$ linearly, then $V$ is called a $G$-\emph{module}.

Let $(G, \,\cdot\,)$ and $(G', \,\circ\,)$ be groups. A map $\phi:G \rightarrow G'$ is a \emph{homomorphism} if $\phi(x \cdot y)=\phi(x) \circ \phi(y)$ for all $x,y \in G$.

The \emph{general linear group} $\text{GL}(n, \mathbb{K})$ is the group of all invertible $n \times n$ matrices over $\mathbb{K}$, whose binary operation is matrix multiplication. 

A \emph{matrix representation} of $G$ is any homomorphism from $G$ into $\text{GL}(n, \mathbb{K})$.

The \emph{trace} of an $n \times n$ matrix $A=(a_{i,j})$ over $\mathbb{K}$, denoted $\text{tr}(A)$, is defined to be $a_{1,1}+ a_{2,2} + \cdots +a_{n,n} \in \mathbb{K}$. 

If $X$ is a matrix representation of $G$, then the \emph{character} of $X$ is a function $\chi:G \rightarrow \mathbb{K}$ defined by $\chi(g)=\text{tr}X(g)$ for any $g \in G$. If $V$ is a $G$-module, then its character, denoted $\chi_V$, is the character of a matrix representation $X$ of $G$ corresponding to $V$.

Let  $S=\{x_1, x_2, \ldots, x_n\}$ and let $V$ be a vector space over $\mathbb{K}$ with basis $\{e_x : x \in S\}$. Let $G$ act on $V$ by $g (\sum{k_xe_x}) = \sum{k_xe_{gx}}$ for $g \in G$ and $k_x \in \mathbb{K}$ by linearly extending the action of $G$ on $S$. The $G$-module $V$ is called the \emph{permutation module} of $G$ on the set $S$.

For instance, consider the permutation module of $\mathfrak{S}_3$ associated with the set $S=\{1, 2, 3\}$. Let $V$ be an 3-dimensional vector space over $\mathbb{K}$ with basis $B=\{e_1, e_2, e_3\}$. We give $V$ an $\mathfrak{S}_3$-module structure by defining $g (\sum{k_xe_x}) = \sum{k_xe_{gx}}$ for any $g \in \mathfrak{S}_3$ and $k_x \in \mathbb{K}$ in an obvious way. For instance, $(1\;2)e_1=e_2,\;(1\;2)e_2=e_1,$ and $(1\;2)e_3=e_3$. The matrix representation $X(g)$ at $g \in \mathfrak{S}_3$ corresponding to $V$ has a 1 in row $i$ and column $j$ if $g \cdot e_j=e_i$, and 0 otherwise.

\begin{center}
$
X(e)=\left(
\begin {matrix}
1 & 0 & 0\\
0 & 1 & 0\\
0 & 0 & 1
\end {matrix}
\right)
,
X((1\,2))=\left(
\begin {matrix}
0 & 1 & 0\\
1 & 0 & 0\\
0 & 0 & 1
\end {matrix}
\right)
,
X((1\,3))=\left(
\begin {matrix}
0 & 0 & 1\\
0 & 1 & 0\\
1 & 0 & 0
\end {matrix}
\right)
,
$
$
X((2\,3))=\left(
\begin {matrix}
1 & 0 & 0\\
0 & 0 & 1\\
0 & 1 & 0
\end {matrix}
\right)
,
X((1\,2\,3))=\left(
\begin {matrix}
0 & 0 & 1\\
1 & 0 & 0\\
0 & 1 & 0
\end {matrix}
\right)
,
X((1\,3\,2))=\left(
\begin {matrix}
0 & 1 & 0\\
0 & 0 & 1\\
1 & 0 & 0
\end {matrix}
\right)
.
$
\end{center}
The reader is encouraged to verify that $X$ is indeed a matrix representation of $\mathfrak{S}_3$, e.g., $X((1\,3\,2))=X((1\,2)(1\,3))=X((1\,2))X((1\,3))$ and $X(e)=X((1\,2)(1\,2))=X((1\,2))X((1\,2))$. We see that the value of the character $\chi_V$ of  $V$ at $g$ equals the number of elements of $B=\{e_1, e_2, e_3\}$ that are fixed by the action of $g \in \mathfrak{S}_3$ in the above example.

Let $H$ be a subgroup of a group $G$. An action of $h \in H$ on the set $G$ defined by $(h,x) \mapsto hxh^{-1}$ is called \emph{conjugation} by $h$. If a group $G$ acts on itself by conjugation, then the set $\{ gxg^{-1}: g \in G \}$ of $x\in G$ is called the \emph{conjugacy class} of $x$. 
Two elements of $\mathfrak{S}_n$ are in the same conjugacy class iff they have the same cycle type.

\begin{thm}[\cite{Dummit2004,Sagan2001}]
\label{thm:conjugacycharacter}
If $K$ is a conjugacy class of $G$, then $g, h \in K$ implies $\chi(g)=\chi(h)$.
\end{thm}

\noindent Theorem~\ref{thm:conjugacycharacter} says that characters are constant on conjugacy classes.

\begin{thm}[\cite{Fulton1991}]
\label{thm:section}
Let $S=\{x_1, x_2,\ldots, x_n\}$ and let $V$ be the associated permutation module of $G$ on $S$. The value of the character $\chi_V$ of $V$ at $g \in G$ equals the number of elements of $S$ that are fixed by the action of $g \in G$.
\end{thm}

The action of $\pi \in \mathfrak{S}_n$ on a Young tableau $t=(t_{i,j})$ of shape $\lambda \vdash n$ is defined here by $\pi t=(\pi(t_{i,j}))$, where $t_{i,j}$ denotes the entry of $t$ in position $(i, j)$. In a similar manner the action of $\pi \in \mathfrak{S}_n$ on tabloids is defined by $\pi\{t\}=\{\pi t\}$. For instance, $(2\;3) \in \mathfrak{S}_3$ acts on a tabloid of shape $\lambda=(2,1)$ as shown below:
\begin{center}
$(2\;3)$
\begin{tabular}{ccc}
\cline{1-2}\noalign{\smallskip}
1 & 2  \\
\cline{1-2}\noalign{\smallskip}
3  \\
\cline{1-1}\noalign{\smallskip}
\end{tabular}
=
\begin{tabular}{ccc}
\cline{1-2}\noalign{\smallskip}
1 & 3  \\
\cline{1-2}\noalign{\smallskip}
2  \\
\cline{1-1}\noalign{\smallskip}
\end{tabular}
\end{center}
We see that $(2\;3) \in \mathfrak{S}_3$ gives a permutation to a tabloid of shape $\lambda$, swapping ``2'' and ``3'' in the tabloid.\\
\indent Suppose $\lambda \vdash n$. Let $V^\lambda$ denote the vector space over the field of real numbers $\mathbb{Re}$ whose basis consists of the set of tabloids of shape $\lambda$, i.e., $V^\lambda=\mathbb{Re}\{\{t_1\},\ldots,\{t_k\}\}$, where $\{t_1\},\ldots,\{t_k\}$ is a complete list of distinct tabloids of shape $\lambda$. Then, $V^\lambda$ is a permutation module of $\mathfrak{S}_n$ on the set of distinct tabloids of shape $\lambda$~\cite{Sagan2001}. 

The following theorem provides a formula to compute the characters of $V^\lambda$ for $\lambda \vdash n$.
\begin{thm}[\cite{Zhao2008}]
\label{thm:permcharacter}
Let $\lambda=(\lambda_1,\ldots,\lambda_i)$ and $\mu=(\mu_1,\ldots,\mu_j)$ be partitions of $n$. The characters of $V^\lambda$ evaluated at an element of $\mathfrak{S}_n$ with cycle type $\mu$ is equal to the coefficient of ${x_1}^{\lambda_1} {x_2}^{\lambda_2}\cdots {x_i}^{\lambda_i}$ in
\begin{center}
$\displaystyle \prod_{k=1}^{j}(x_1^{\mu_k}+x_2^{\mu_k}+\cdots+x_i^{\mu_k}).$
\end{center}
\end{thm}

We now consider task reassignments by means of a group action for $n$-task-$n$-agent assignments represented by the set of standard assignment tabloids of shape $\lambda \vdash n$. We define the action of $\pi \in \mathfrak{S}_n$ on standard assignment tabloids in exactly the same way as above.  

\begin{figure}[h!]
  \centering
$\{S_{\mu}^{1}\}=\begin{tabular}{ccc}
\cline{1-3}\noalign{\smallskip}
1 & 2  & 3\\
\cline{1-3}\noalign{\smallskip}
4  \\
\cline{1-1}\noalign{\smallskip}
\end{tabular}
,
\{S_{\mu}^{2}\}=\begin{tabular}{ccc}
\cline{1-3}\noalign{\smallskip}
1 & 2  & 4\\
\cline{1-3}\noalign{\smallskip}
3  \\
\cline{1-1}\noalign{\smallskip}
\end{tabular}
,
\{S_{\mu}^{3}\}=\begin{tabular}{ccc}
\cline{1-3}\noalign{\smallskip}
1 & 3  & 4\\
\cline{1-3}\noalign{\smallskip}
2  \\
\cline{1-1}\noalign{\smallskip}
\end{tabular}
,
\{S_{\mu}^{4}\}=\begin{tabular}{ccc}
\cline{1-3}\noalign{\smallskip}
2 & 3  & 4\\
\cline{1-3}\noalign{\smallskip}
1  \\
\cline{1-1}\noalign{\smallskip}
\end{tabular}
$\,.
\caption{Standard assignment tabloids of shape $\mu=(3, 1)$.}
\label{fig:StandardAssignmentTabloids}
\end{figure}

Consider $g = (1\,2) \in \mathfrak{S}_4$ acts on each standard assignment tabloid $\{S_{\mu}^{k}\}$ for $1\leq k \leq 4$. For instance, $(1\,2)\{S_{\mu}^{3}\}$ swaps task 1 and task 2 in the $n$-task-$n$-agent assignments represented by $\{S_{\mu}^{3}\}$. We see that $g\{S_{\mu}^{1}\}=\{S_{\mu}^{1}\},\; g\{S_{\mu}^{2}\}=\{S_{\mu}^{2}\}$, $g\{S_{\mu}^{3}\}=\{S_{\mu}^{4}\}$, and $g\{S_{\mu}^{4}\}=\{S_{\mu}^{3}\}$. Note that swapping elements in the same row of the given tabloid remains the tabloid invariant. We see that $g$ fixes $\{S_{\mu}^{1}\}$ and $\{S_{\mu}^{2}\}$ only. The following proposition is the main result of this section.

\begin{prop}
\label{prop:section}
The number of standard assignment tabloids of shape $\lambda \vdash n$ fixed by the (task reassignment) action of $g \in \mathfrak{S}_n$ is the character of $V^\lambda$ evaluated on the conjugacy class of $g \in \mathfrak{S}_n$.
\end{prop}
\begin{proof}
Let $\mathfrak{S}_n$ act on the set $S$ consisting of a complete list of distinct standard assignment tabloids of shape $\lambda \vdash n$ as above. Then, we see that its associated permutation module of $\mathfrak{S}_n$ on $S$ is simply $V^\lambda$. The number of standard assignment tabloids of shape $\lambda \vdash n$ fixed by the (task reassignment) action of $g \in \mathfrak{S}_n$ is the character of $V^\lambda$ at $g \in \mathfrak{S}_n$ by Theorem~\ref{thm:section}. Since characters are constant on conjugacy classes by Theorem~\ref{thm:conjugacycharacter}, the number of standard assignment tabloids of shape $\lambda \vdash n$ fixed by the (task reassignment) action of $g \in \mathfrak{S}_n$ is the character of $V^\lambda$ evaluated on the conjugacy class of $g\in \mathfrak{S}_n$.
\end{proof}
By Proposition~\ref{prop:section}, we see how a task reassignment represented by $g \in \mathfrak{S}_n$ affects the elements of the search space of the $n$-task-$n$-agent assignment problem represented by the set of standard assignment tabloids of shape $\lambda \vdash n$. Using Lemma~\ref{lem:Numberofstandardassignmenttabloids}, the size of the search space represented by the set of standard assignment tabloids of a given shape $\lambda \vdash n$ is obtained. By subtracting the number in Proposition~\ref{prop:section} from the above size of the search space, we obtain the total number of standard assignment tabloids of a given shape $\lambda \vdash n$ that is not invariant by a task reassignment represented by $g \in \mathfrak{S}_n$. The higher the total number, the more elements of the search space become affected and the more computations for task reassignments are required as a result. Note that by Proposition~\ref{prop:EquivalentStandardAssignmentsTableaux}, a standard assignment tabloid $\{S_\lambda\}$ fixed by the action of $g \in \mathfrak{S}_n$ has the same task turnaround time before and after the action of $g \in \mathfrak{S}_n$, which does not need a task reassignment by $g \in \mathfrak{S}_n$ at all.

\begin{table}[h!]
\centering
\caption{Characters of $V^\lambda$ for $\mathfrak{S}_4$ ($\phi^\lambda$: the character of $V^\lambda$, $K_\mu$: the conjugacy class of $\mathfrak{S}_4$ with cycle type $\mu$)~\cite{Zhao2008}.\\}
\begin{tabular}{c|c|c|c|c|c}

&$K_{(1,1,1,1)}$ & $K_{(2,1,1)}$ & $K_{(2,2)}$ & $K_{(3,1)}$ & $K_{(4)}$\\
\hline
$\phi^{(1,1,1,1)}$ & 24 & 0 & 0 & 0 & 0 \\
$\phi^{(2,1,1)}$ & 12 & 2 & 0 & 0 & 0 \\
$\phi^{(2,2)}$ & 6 & 2 & 2 & 0 & 0 \\
$\phi^{(3,1)}$ & 4 & 2 & 0 & 1 & 0 \\
$\phi^{(4)}$& 1 & 1 & 1 & 1 & 1 \\
\end{tabular}
\label{table:CharacterOfVectorSpace}
\end{table}


Now, consider Table~\ref{table:CharacterOfVectorSpace}, which shows the list of characters of $V^\lambda$ for each conjugacy class of $\mathfrak{S}_4$~\cite{Zhao2008}. The characters of $V^\lambda$ in Table~\ref{table:CharacterOfVectorSpace} can also be computed by means of Theorem~\ref{thm:permcharacter}. For instance, we obtain $\phi^{(2, 2)}$ at $K_{(2, 1, 1)}$ in Table~\ref{table:CharacterOfVectorSpace} by computing the coefficient of $x_1^2x_2^2$ in $(x_1^2+x_2^2)(x_1+x_2)^2$, which is 2. (The interested reader may refer to~\cite{Zhao2008, Sagan2001} for further details.) Note that the column of $K_{(1,1,1,1)}$ in Table~\ref{table:CharacterOfVectorSpace} corresponds to the dimension of $V^\lambda$, which indicates the number of distinct standard assignment tabloids of shape $\lambda \vdash n$ for $n = 4$. Since $\phi^{(3,1)}$ at $K_{(2,2)}$ is 0 in Table~\ref{table:CharacterOfVectorSpace}, we see that in Figure~\ref{fig:StandardAssignmentTabloids} the number of standard assignment tabloids of shape $\lambda=(3, 1)$ fixed by the (task reassignment) action of $(1\,2)(3\,4) \in \mathfrak{S}_4$ is 0 by Proposition \ref{prop:section}, where $(1\,2)(3\,4) \in \mathfrak{S}_4$ has cycle type $(2, 2)$.

\section{Conclusions}
\label{sec:Conclusions}
This paper presented a framework for representing task assignments and reassignments in distributed systems using Young tableaux and finite groups focused on finite symmetric groups. We showed that a task assignment with a partition is naturally represented by a Young tableau and that task reassignments are described by means of a group action on a set of Young tableaux. \\
\indent We introduced a standard assignment tableau (respectively, a standard generalized assignment tableau) to represent an $n$-task-$n$-agent assignment (respectively, an $n$-task-$m$-agent assignment ($n > m$)) in a visual and compact manner, while representing a logical partition of agents (respectively, tasks) in a distributed system. \\
\indent We discussed row-equivalence classes of Young tableaux to represent certain equivalence classes of standard assignment tableaux. Then, we discussed the search space reduction using equivalence classes of standard assignment tableaux and showed how task reassignments by means of a group action affect the elements of the search space for the $n$-task-$n$-agent assignment problem. We also discussed an equivalence relation and a total order on a set of standard generalized assignment tableaux of a given shape in order to explore the selected search space for the $n$-task-$m$-agent assignment problem ($n > m$) in a well-defined and systematic manner by means of group-theoretical methods.\\
\indent By raising the expressiveness of task assignments, our approach is able to employ some of the known results of Young tableaux and group theory for task assignments and reassignments in distributed systems.\\
\indent There are a wide variety of tableaux (e.g.  \emph{skew tableaux}~\cite{Sagan2001}, \emph{oscillating tableaux}~\cite{Stanley1997}, etc.) and their algorithms (e.g. \emph{row insertion}~\cite{Sagan2001}, \emph{deletion}, \emph{forward and backward slides}~\cite{Stanley1997}, etc.) used in combinatorics and group theory, which have not been discussed in this paper. Variants of assignment tableaux can be considered by means of those tableaux. When applying our approach to certain types of distributed systems, one may need to link additional constraints (e.g. task or node priorities, task dependencies~\cite{El-Rewini1998}, etc.) or data structure to entries or cells in an assignment tableau. One may also need to consider the variants of assignment tableaux to represent the specific kinds of tasks or nodes in certain types of distributed systems. We leave it to future work to consider the possible variants of assignment tableaux along with their algorithms in distributed systems and to examine them from both theoretical and practical perspectives.

\nocite{*}
\bibliographystyle{gCOM}
\bibliography{dkim}

\section*{Appendix A. Young tableaux for job-shop and flow-shop scheduling}
There are a wide variety of task assignment and scheduling problems, such as \emph{job-shop}~\cite{Yamada1997,Seda2007}, \emph{flow-shop shop}~\cite{Kumar2014, Seda2007}, and \emph{resource-constrained project scheduling problems}~\cite{Hartmann2010}. Job-shop and flow-shop scheduling have also been discussed in a distributed system environment~\cite{Tilborg1991, Bettati1992}. In this appendix we briefly consider assignment objects using Young tableaux for the job-shop and flow-shop scheduling problem.\\
\indent The job-shop scheduling problem~\cite{Yamada1997,Seda2007}, in its classical form, can be described by a set of $n$ jobs $\{J_i\}_{1\leq i \leq n}$ that is to be processed on a set of $m$ machines (or agents) $\{M_j\}_{1\leq j \leq m}$. Each job consists of a chain of $m$ operations and has a machine order to be processed, in which $m$ operations of job $J_i$ are processed on $m$ different machines. The processing of job $J_i$ on  machine $M_j$ is denoted by operation $O_{ij}$. Operation $O_{ij}$ requires the processing time $p_{ij}$ for the use of machine $M_j$ in an uninterruptable manner. The scheduling problem is to find an assignment of all operations to minimize the \emph{makespan} which is the maximum of the completion times (or finishing times) of all jobs (see~\cite{Yamada1997,Seda2007} for detailed assumptions and constraints for the job-shop scheduling problem).\\
\indent Meanwhile, in the flow-shop scheduling problem~\cite{Kumar2014, Seda2007}, each job consists of a chain of $m$ operations and has the exact same machine order to be processed, in which $m$ operations of a job are processed on $m$ different machines. The scheduling problem here is to find the job sequences on the machines that minimize the makespan~\cite{Seda2007}. The flow-shop scheduling problem is a special case of the job-shop scheduling problem~\cite{Kumar2014}.\\
\indent Now, consider the following machine orders of the jobs: \\
\\
$J_1:M_4 \rightarrow M_3 \rightarrow M_1 \rightarrow M_2$,\\
$J_2:M_3 \rightarrow M_2 \rightarrow M_1 \rightarrow M_4$,\\
$J_3:M_2 \rightarrow M_3 \rightarrow M_4 \rightarrow M_1$.\\

Note that the machine orders of the jobs for the classical, deterministic job-shop scheduling and flow-shop scheduling problem are given in advance, while job orders on machines may vary by different schedules. Note also that in the flow-shop scheduling problem, $J_1$, $J_2$, and $J_3$ have to be the same. An example of job orders on machines is as follows: \\

\noindent $M_1:J_1 \rightarrow J_2 \rightarrow J_3$,\\
$M_2:J_3 \rightarrow J_2 \rightarrow J_1$,\\
$M_3:J_2 \rightarrow J_1 \rightarrow J_3$,\\
$M_4:J_1 \rightarrow J_3 \rightarrow J_2$.\\

The assignment tableaux we have discussed in the previous sections cannot describe the job orders on machines. One possible way of describing the above job assignment using a generalized Young tableau is as follows:
$\\$
\begin{center}
$\;\;\Yboxdim13pt\young(123,321,213,132)$\;\;\;
\end{center}
$\\$
The rows in the above generalized Young tableau describes the machines, in which the first row describes the first machine (i.e., $M_1$), the second row describes the second machine (i.e., $M_2$), and so on. The columns in the above generalized Young tableau describes the job orders, in which the first column describes (the index of) the first job in the job orders, the second column describes (the index of) the second job in the job orders, and so on. 
For a non-classical job-shop scheduling problem, each job may consist of a chain of the different number of operations. In this case we may need to use empty cells in a tableau to describe the different number of job orders on machines. \\
\indent To define assignment objects using Young tableaux and groups in other types of scheduling problems, such as resource-constrained project scheduling problems~\cite{Hartmann2010}, are problem-specific and have not been well-established thus far, which requires future research.

\section*{Appendix B. Computational tools}
In Section~\ref{subsec:generalized} we described an equivalence relation $\sim_{tr}$ for task reassignments on $X_n^\lambda=\{t_1, t_2, \ldots,t_n\}$ which is a set of standard generalized assignment tableaux $t_i (1\leq i \leq n)$ of shape $\lambda$. The following figure describes how the GAP~\cite{GAP2014, Holt2005, Hulpke2010} computer algebra system can be used for finding the size of the equivalence class of $t \in X_n^\lambda$ with respect to the equivalence relation $\sim_{tr}$ by regarding $X_n^\lambda$ as a set $\{1, 2, \ldots, n\}$ and defining the action of $G\leq \mathfrak{S}_n$ on $\{1, 2, \ldots, n\}$ simply by $gi=g(i)$ for $g \in G$. 
\begin{figure}[ht!]
\small
\begin{framed}
1:\;\;gap$>$ g:=Group((1,2,3,4),(47,48,49,50));\\
2:\;\;\emph{$<$permutation group with 2 generators$>$;}\\
3:\;\;gap$>$ Elements(g);\\
4:\;\;\emph{[ (), (47,48,49,50), (47,49)(48,50), (47,50,49,48), (1,2,3,4), (1,2,3,4)(47,48,49,50),} \\
5:\;\;\emph{(1,2,3,4)(47,49)(48,50), (1,2,3,4)(47,50,49,48), (1,3)(2,4), (1,3)(2,4)(47,48,49,50),}\\
6:\;\;\emph{(1,3)(2,4)(47,49)(48,50), (1,3)(2,4)(47,50,49,48),} \\
7:\;\;\emph{(1,4,3,2), (1,4,3,2)(47,48,49,50), (1,4,3,2)(47,49)(48,50), (1,4,3,2)(47,50,49,48) ]}\\
8: gap$>$ Order(g);\\
9: \emph{16}\\
10:\;\;gap$>$ orbits:=Orbits(g, [1..50]);\\
11:\;\;\emph{[ [1, 2, 3, 4], [5], [6], [7], [8], [9], [10], [11], [12], [13], [14], [15], [16], [17], [18], [19],}\\ 
12:\;\;\emph{[20], [21], [22], [23], [24], [25], [26], [27], [28], [29],[30], [31], [32], [33], [34], [35], [36], }\\
13:\;\;\emph{[37], [38], [39], [40], [41], [42], [43], [44], [45], [46], [47, 48, 49, 50] ]}\\
14:\;\;gap$>$ OrbitLength(g,48);\\
15:\;\;\emph{4}\\
16:\;\;gap$>$ Size(orbits);\\
17:\;\;\emph{44}
\end{framed}
\caption{Examples of Propositions~\ref{prop:orbit} and~\ref{prop:sizeoforbit} using GAP.}
\label{fig:GAP}
\end{figure}

Lines 1--2 in Figure~\ref{fig:GAP} indicates that we choose $G \leq \mathfrak{S}_n$ that is generated by the cycles (1\;2\;3\;4) and (47\;48\;49\;50). Lines 3--7 in Figure~\ref{fig:GAP} shows the elements of $G$. Lines 10--13 in Figure~\ref{fig:GAP} show all the orbits when $G$ acts on a set $\{1, 2, \ldots, 50\}$ by $gi=g(i)$ for $g \in G$. Using the renaming map $i \mapsto t_i$, we may interpret them as all the equivalence classes of $X_{50}^\lambda=\{t_1, t_2, \ldots, t_{50}\}$ with respect to $\sim_{tr}$ when $G$ acts on $X_{50}^\lambda$ by $gt_i=t_{g(i)}$ for $g \in G$. For instance, [1, 2, 3, 4] in line 11 can be interpreted as $[t_1, t_2, t_3, t_4]$. Lines 14--15 in Figure~\ref{fig:GAP} indicates that the size of the orbit of $48 \in \{1, 2, \ldots, 50\}$ is 4. We may interpret it as the size of the equivalence class of $t_{48} \in X_n^\lambda$ is 4 with respect to $\sim_{tr}$. Finally, lines 16--17 show that the number of orbits is 44. We may interpret it as the number of equivalence classes of $X_{50}^\lambda$ with respect to $\sim_{tr}$ is 44.

We also provide a simple tool written in GNU C++\cite{GNU} for the $n$-task-$n$-agent assignment problem using \emph{genetic algorithms}~\cite{Sahu2007} and Young tableaux. Metaheuristic methods, such as genetic algorithms, are often used for task assignment problems, since finding an optimal task assignment is often intractable~\cite{El-Rewini1998, Sinnen2007}. The purpose of our tool\footnote{Source codes and sample data are available at http://www.airesearch.kr/downloads/STG.zip.}, called simple tableaux-based genetic algorithms (STG), is to show how Young tableaux and their partitions are applied to the existing genetic algorithms for task assignment problems. The following table shows an example of (file) input and output of STG for the $n$-task-$n$-agent assignment problem for $n=20$.
\\
\begin{longtable}{l}
\caption{An example of (file) input and output of STG for the $n$-task-$n$-agent-assignment problem for $n=20$.}
\label{table:STG}\\
\hline\\
\small 1: \textbf{Input:}\\\\
\small 2: NumberOfTasks 20\\\\
\small 3: NumberOfAgents 20\\\\
\small 4: $\cdots$\\
\\
\small 5: Task$\textunderscore$Length 10 20 30 40 50 60 70 80 90 100 110 120 130 140 150 160 170 180 190 200\\\\
\small 6: Processing$\textunderscore$Capacity 10 20 30 40 50 60 70 80 90 100 110 120 130 140 150 160 170 180 190 200\;\\\\
\small 7: $\cdots$\\
\\
\small 8: \#GAs \\\\
\small 9: CrossoverRate   0.8\\\\
\small 10: MutationRate    0.005\\\\
\small 11: $\cdots$
\\\\
\small 12: \#Young Tableau : Y, Young Tabloid : N\\\\
\small 13: YoungTableau    Y\\\\
\small 14: NumberOfPartitions  4\\\\
\small 15: TableauShape        6 6 4 4\\\\
\small 16: $\cdots$\\
\\
\small 17: 1\\\\
\small 18: 2\\\\
\small 19: $\cdots$\\
\\
\small 20: 19\;\;\;16\;\;\;17\\\\
\small 21: 20\;\;\;18\;\;\;19\\\\
\\\\
\small 22: \textbf{Output:}\\\\
\small 23: (GENERATION 800)\\\\
\small 24: ********************** Candidate Solution********************\\\\
\small 25: Candidate Solution Fitness:0.205212\\\\
\small 26: 1 2 3 4 5 6 7 9 8 11 12 13 14 16 10 17 18 15 19 20\\\\
\small 27: $\cdots$\\\\
\small 28: Tableau Shape: 6 6 4 4\\\\
\small 29: $\cdots$\\\\
\small 30: $\{\{1,2,3,4,5,6\},\{7,9,8,11,12,13\},\{14,16,10,17\},\{18,15,19,20\}\}$,\\\\
\small 31: $\cdots$\\
\\
\hline\\
\end{longtable}

The input of STG specifies the number of tasks, number of agents, the list of task lengths, the list of processing capacities of agents, parameters of genetic algorithms, etc. The first entry of the task length list (see line 5) in Table~\ref{table:STG} corresponds to the task length of task ID 1, the second entry corresponds to the task length of task ID 2, and so on. Similarly, the first entry of the processing capacity list (see line 6) in Table~\ref{table:STG} corresponds to the processing capacity of agent ID 1, the second entry corresponds to the processing capacity of agent ID 2, and so on. At the bottom of ``Input'' in Table~\ref{table:STG} (see lines 17--21), the leftmost numbers denote task IDs (i.e., task IDs 1, 2, $\ldots$, 19, and 20), while their following numbers denote their precedence lists. For instance, task IDs 1 and 2 have no precedence list, while task ID 19 (respectively, task ID 20) has its precedence list [16, 17] (respectively, [18, 19]). \\
\indent The output of STG shows the candidate solution represented by the one-line permutation notation for the $n$-task-$n$-agent assignment problem after a certain number of generations (see line 26 in Table~\ref{table:STG}). Then, the candidate solution represented by the one-line permutation notation is converted into the corresponding standard assignment tableau based on the given tableau shape. Finally, the converted candidate solution $\{\{1,2,3,4,5,6\},\{7,9,8,11,12,13\},\{14,16,10,17\},\{18,15,19,20\}\}$ (see line 30 in Table~\ref{table:STG}) represents the standard assignment tableau of shape $(6, 6, 4, 4) \vdash 20$, where the entries of its first row (reading from left to right) are 1, 2, 3, 4, 5, 6; the entries of its second row 7, 9, 8, 11, 12, 13; the entries of its third row 14, 16, 10, 17; and the entries of its last row are 18, 15, 19, 20.
\end{document}